\documentclass{jfm}

\usepackage{graphicx}
\usepackage{natbib}
\usepackage[utf8]{inputenc}
\usepackage[T1]{fontenc}
\usepackage{graphicx}
\usepackage{amsmath}
\usepackage{amsfonts}
\usepackage{amssymb}
\usepackage{caption}
\usepackage{subcaption}
\usepackage{epstopdf}
\usepackage{authblk}

\DeclareGraphicsExtensions{.pdf,.png,.eps}

\ifCUPmtlplainloaded \else
  \checkfont{eurm10}
  \iffontfound
    \IfFileExists{upmath.sty}
      {\typeout{^^JFound AMS Euler Roman fonts on the system,
                   using the 'upmath' package.^^J}%
       \usepackage{upmath}}
      {\typeout{^^JFound AMS Euler Roman fonts on the system, but you
                   dont seem to have the}%
       \typeout{'upmath' package installed. JFM.cls can take advantage
                 of these fonts,^^Jif you use 'upmath' package.^^J}%
       \providecommand\upi{\pi}%
      }
  \else
    \providecommand\upi{\pi}%
  \fi
\fi
\newcommand{\pderiv}[2]{\frac{\partial #1}{\partial #2}}

\ifCUPmtlplainloaded \else
  \checkfont{msam10}
  \iffontfound
    \IfFileExists{amssymb.sty}
      {\typeout{^^JFound AMS Symbol fonts on the system, using the
                'amssymb' package.^^J}%
       \usepackage{amssymb}%
         \let\leq=\leqslant
         \let\geq=\geqslant
      }{}
  \fi
\fi

\ifCUPmtlplainloaded \else
  \IfFileExists{amsbsy.sty}
    {\typeout{^^JFound the 'amsbsy' package on the system, using it.^^J}%
     \usepackage{amsbsy}}
    {\providecommand\boldsymbol[1]{\mbox{\boldmath $##1$}}}
\fi

\renewcommand{\vec}[1]{\boldsymbol{#1}}

\newcommand{\xx}{\vec{x}}

\newsavebox{\astrutbox}
\sbox{\astrutbox}{\rule[-5pt]{0pt}{20pt}}

\newcommand{\e}{\textup{e}}

\title[Leading-Edge Serrations]{An Analytic Solution for the Noise Generated by Gust-Aerofoil Interaction for Plates with Serrated Leading Edges}

\author[1]{L\ls O\ls R\ls N\ls A\ns J.\ns A\ls Y\ls T\ls O\ls N
 \thanks{Email address for correspondence: L.J.Ayton@damtp.cam.ac.uk}}
 \affil[1]{Department of Applied Mathematics and Theoretical Physics, University of Cambridge,
Wilberforce Road, CB3 0WA, UK}
\author[2]{J\ls A\ls E\ns W\ls O\ls O\ls K\ns K\ls I\ls M
} 
  \affil[2]{Aerodynamics \& Flight Mechanics Research Group, University Road, University of Southampton, Southampton, SO17 1BJ, UK}

\begin{document}

\maketitle

\begin{abstract}
This paper presents an analytic solution for the sound generated by an unsteady gust interacting with a semi-infinite flat plate with a serrated leading edge in a background steady uniform flow. Viscous and non-linear effects are neglected. The Wiener-Hopf method is used in conjunction with a non-orthogonal coordinate transformation and separation of variables to permit analytical progress. The solution is obtained in terms of a modal expansion in the spanwise coordinate, however for low- and mid-range incident frequencies only the zeroth order mode is seen to contribute to the far-field acoustics, therefore the far-field noise can be quickly evaluated.
The solution gives insight into the potential mechanisms behind the reduction of noise for plates with serrated leading edges compared to those with straight edges, and predicts a logarithmic dependence between the tip-to-root serration height and the decrease of far-field noise. The two mechanisms behind the noise reduction are proposed to be an increased destructive interference in the far field, and a redistribution of acoustic energy from low cuton modes to higher cutoff modes as the tip-to-root serration height is increased.
The analytic results show good agreement in comparison with experimental measurements. The results are then compared against numerical predictions for the sound generated by a spanwise invariant line vortex interacting with a flat plate with serrated leading edge. Good agreement is also seen between the analytical and numerical results as frequency and tip-to-root ratio are varied.
 \end{abstract}

\section{Introduction}\label{sec:intro}
Leading-edge noise is generated by the unsteady wakes of a forward rotor row impinging on a rearward stator row within an aeroengine. It is well known as a dominant source of aircraft noise \citep{Parry} and as such has sparked a large amount of research aimed at understanding and controlling noise levels \citep{Amiet,Atassi,MK1995,Lock,AytonChai}. Recent interest in silent owl flight has led to research in a number of aerofoil adaptations as a way to reduce aerofoil-turbulence interaction noise as discussed by \citet{Lilley}. The leading-edge comb \citep{Graham} appears as a serration to the leading edge of the wing, and through experimental \citep{GeyerAIAA2016,Joseph1} and numerical investigations \citep{Haeri,JaeWook} has been seen as an effective way to reduce leading-edge noise. Despite these results, it is still not fully understood why the serrated edge is such an effective way to reduce leading-edge noise, therefore analytic solutions for simple leading-edge interaction noise models are sought to illuminate the physical noise-reduction mechanisms within the flow, and hence aid in designing optimal leading-edge geometries for silent blade operation.

Current analytic models for leading-edge serrations such as the gust-interaction noise model of \citet{LyuLE} or the sound scattering model of \citet{Huang} typically rely on standard Fourier series expansions in the spanwise coordinate (along which the serration lies) and numerical techniques to eventually solve the final governing equations. A downside of this is that it becomes difficult to extract precise information from the solutions as to why a reduction of noise is possible, as one cannot pick apart the final solution to determine when and where each term has come from. 
Solutions from the semi-analytic iterative method using the Schwartzchild technique \citep{LyuLE} predict that the noise reduction for rapidly serrated edges may differ from those predicted experimentally \citep{Joseph1} since it is typical that experimentally noise is measured in a restricted arc centred above the plate. This misses the far upstream and downstream directions; the semi-analytic solution \citep{LyuLE} predicts that the acoustic directivity pattern is significantly distorted by rapid serrations, which could result in larger or smaller pressure magnitudes at shallow angles than the effects seen directly above the plate (at $\theta=90^{\circ}$).

Both \cite{LyuLE}'s analytical and \citet{Joseph1}'s experimental results suggest that phase interference in the scattered field are key to noise reduction for rapidly serrated edges, and a greater noise reduction is possible for more rapidly serrated edges due to an increased destructive interference. However if the far field were to experience a region of destructive interference it would be natural to assume there could be a corresponding region of constructive interference. The analytic solution obtained via Green's functions for the sound generated by individual vortices interacting with a wavy leading edge by \citet{JamesM} indicates certain leading-edge profiles do increase noise as opposed to reducing it. Key to an increase or decrease in noise is the relative angle between the vortex path and the leading edge, which is also an important factor for swept leading edges \citep{sweptLE}. 
\citet{Huang} who uses a Wiener-Hopf approach proposes a reduction of far-field noise due to a cutoff of the scattered frequency, as is the case for the swept edge \citep{sweptLE}, however similarly to \citet{LyuLE}, it is difficult to infer this conclusion directly from the mathematics as the details are unfortunately hidden by the complexity surrounding the Fourier series expansions and numerical Wiener-Hopf factorisation.

This paper attempts to provide a simpler analytic solution for the noise generated by a serrated leading edge, which can be used to understand the mechanisms allowing for noise reduction. We avoid the need for any numerical steps during the calculation of the far-field pressure by using a sequence of variable transformations to convert the governing equation and boundary conditions into a form suitable for solving analytically using the Wiener-Hopf technique. We also avoid the need for numerically factorising the Wiener-Hopf kernel. This approach follows the work of \citet{Envia} who considered the effects of blade sweep in a finite-span channel with rigid walls.

The layout of this paper is as follows. The governing equation and boundary conditions for the problem are given in Section \ref{sec:formulation} along with the required transformation of coordinates to allow for a simple solution. The solution is found in Section \ref{sec:solution} using separation of variables and the Wiener-Hopf technique. Section \ref{sec:results} contains results for the scattered acoustic far field which we compare to numerical simulations adapted from \citet{Turner2017}. A brief discussion of the adapted numerical method used is given in Section \ref{sec:numerics}. We discuss the conclusions of this paper in Section \ref{sec:conc}.

\section{Formulation of the Problem}\label{sec:formulation}

\begin{figure}
 \centering
 \includegraphics[width=0.5\textwidth, clip, trim=0.3cm 0.5cm 0cm 0.2cm]{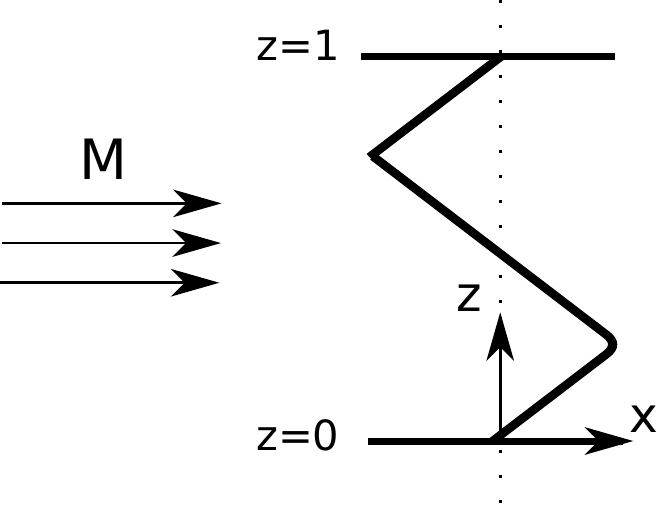}
\caption{Uniform steady flow in the $x$ direction with unsteady convective gust perturbation interacts with a rigid plate $y=0$ with serrated edge $x=c F(z)$.}
\label{fig:serrationchannel}
\end{figure}

We consider the interaction of a convective unsteady gust in uniform flow of Mach number $M$ 
over a semi-infinite flat plate with a serrated leading edge. To simplify the problem we consider a single wavelength of the serration, thus the blade lies in the region $y=0$, $x>c F(z)$, $0\leq z\leq1$ as depicted in Figure \ref{fig:serrationchannel}. Here $c$ is a positive real constant parameterising the so-called `tip-to-root' height of the serration (which is given by $c/2$). We non-dimensionalise lengths by the  wavelength of the serration, and velocities by the far upstream steady velocity. Pressure is non-dimensionalised with respect to the far upstream steady density and velocity.

We suppose the serration is single-frequency, therefore define
\begin{equation}
 F(z)=
\begin{cases}
 z, \qquad\qquad &z\in[0,\frac{1}{4})\\
\frac{1}{2}-z, \qquad &z\in(\frac{1}{4},\frac{3}{4})\\
z-1, \qquad &z\in(\frac{3}{4},1]
\end{cases}
\label{eq:F}
\end{equation}
Note the channel boundaries are located at serration midpoints to ensure any effects of the sharp tip and root at $z=1/4,3/4$ are fully accounted for and the edges do not interfere with these key features.

The unsteady gust incident from far upstream takes the form
\begin{equation}
\vec{v}_{g}=\vec{A}\e^{i k_{1}x+i k_{2}y+i k_{3}z-i \omega t},
\end{equation}
where the amplitude, $\vec{A}=(A_{1},A_{2},A_{3})^{T}$, is constant. 

We consider two cases for the scattered field: a) the channel has rigid walls thus there is a single wavelength serrated blade in a duct; and b) the channel has periodic boundary conditions.
In both cases we set $A_{3}=0$, and as we are dealing with an infinitely thin plate the only amplitude term from the gust present in our problem will be $A_{2}$, which for simplicity we set to unity. Case a) allows us to compare the effects of a serrated edge against that of a swept edge in a channel \citep{Envia}, whilst case b) allows us to consider a spanwise-infinite plate with a periodic serrated leading edge.

We decompose the unsteady flow field into a convective gust part and an acoustic response part, $\vec{v}=\vec{v}_{g}+\vec{v}_{a}$, and write the response as $\vec{v}_{a}=\nabla G$. We suppose $G$ is harmonic in time $\sim\e^{-i \omega t}$ therefore spatially satisfies the convected Helmholtz equation,
\begin{equation}
\beta^{2}\pderiv{^{2}G}{x^{2}}+\pderiv{^{2}G}{y^{2}}+\pderiv{^{2}G}{z^{2}}+2i kM\pderiv{G}{x}+k^{2}G=0,
\label{eq:gov}
\end{equation}
where $\beta^{2}=1-M^{2}$ and $k=\omega/ c_{0}$ with $c_{0}$ the speed of sound of the background steady flow. Since the gust convects with the background flow, we require $k=k_{1}M$.
The zero normal velocity boundary condition on the aerofoil surface requires
\begin{subequations}
\begin{equation}
\left.\pderiv{G}{y}\right|_{y=0}=-\e^{i k_{1} x+i k_{3}z}\qquad x> c F(z).
\end{equation}
We also impose continuity of the potential upstream
\begin{equation}
\left.\Delta G\right|_{y=0}=0\qquad x<c F(z).
\end{equation}
\end{subequations}
Finally, in case a) the rigid channel walls require
\begin{subequations}
\begin{equation}
 \left.\pderiv{G}{z}\right|_{z=0,1}=0,
\end{equation}
or case b) the periodic conditions yield
\begin{equation}
 G|_{z=0}=G|_{z=1}\e^{i \alpha k_{3}},\qquad  \left.\pderiv{G}{z}\right|_{z=0}=\left.\pderiv{G}{z}\right|_{z=1}\e^{i \alpha k_{3}},
\end{equation}
\end{subequations}
where $\alpha$ is a real constant enforcing the level of periodicity of the problem, e.g. if $\alpha=-1$ the solution must be periodic over one serration wavelength or if $\alpha=-1/2$ the solution must be periodic over two serration wavelengths. We will restrict results in this paper to $\alpha=-1$ however retain the parameter in our calculations to allow a future study on varying the periodicity of the solution.

To simplify the governing equation, \eqref{eq:gov}, we apply a convective transform,
\begin{equation}
h=G(x,y,z)\e^{i k_{1} M^{2}x/\beta^{2}},
\label{eq:conv}
\end{equation}
to eliminate the convective terms. The resulting governing equation and boundary conditions for $h(x,y,z)$ are
\begin{subequations}
\begin{equation}
\beta^{2}\pderiv{^{2}h}{x^{2}}+\pderiv{^{2}h}{y^{2}}+\pderiv{^{2}h}{z^{2}}+\left(\frac{k_{1}M}{\beta}\right)^{2}h=0,\label{eq:hgov}
\end{equation}
\begin{equation}
\left.\pderiv{h}{y}\right|_{y=0}=-\e^{i \frac{k_{1}}{\beta^{2}} x+i k_{3}z}\qquad x> c F(z),
\end{equation}
\begin{equation}
\left.\Delta h\right|_{y=0}=0\qquad x<c F(z),
\end{equation}
\label{eq:WHgov}
\end{subequations}
accompanied by the option of a rigid vertical wall condition (case a) or periodic condition (case b) on $z=0,1$ as before.

This set of equations, \eqref{eq:WHgov}, forms a mixed boundary condition problem in regions $x\gtrless cF(z)$ therefore we wish to employ the Wiener-Hopf technique. However to make this problem more amenable to the Wiener-Hopf technique we wish to shift the two regions to some $\xi\gtrless0$ making them independent of the spanwise variable. To do so we perform the following transformation of coordinates (adapted from the transformation used by \citet{Envia} for a swept edge);
\begin{subequations}
 \begin{align}
  \xi&=\frac{\sqrt{1-\gamma^{2}}}{\beta}x-\gamma F(z),
\\
\eta&= y,
\\
\zeta&= z,
\\
\gamma&=\frac{c}{\sqrt{\beta^{2}+c^{2}}}.
 \end{align}
\end{subequations}

The transformed governing equation becomes
\begin{subequations}
 \begin{equation}
  \nabla^{2}_{\xi,\eta,\zeta}h-2\gamma F'(\zeta)\pderiv{^{2}h}{\xi\partial\zeta}+(d M)^{2}h=\gamma \left(\delta(\zeta-\frac{3}{4})-\delta(\zeta-\frac{1}{4})\right)\pderiv{h}{\xi},
 \end{equation}
where we set $d=k_{1}/\beta$ and use $\delta(x)$ to denote the Dirac delta function arising due to the discontinuities in $F'(\zeta)$ at the peaks and roots of the serration. Note the right hand side equals $\gamma F''(\zeta)\pderiv{h}{\xi}$, and derivatives of $F$ are formally weak derivatives.

The boundary conditions become
\begin{equation}
 \left.\pderiv{h}{\eta}\right|_{\eta=0}=-\e^{i\kappa\xi+i k_{3}\zeta}\e^{i\kappa\gamma F(\zeta)}\quad \xi>0,\label{eq:zerovel}
\end{equation}
where $\kappa=d/\sqrt{1-\gamma^{2}}$,
\begin{equation}
 \Delta h|_{\eta=0} = 0\quad \xi<0,\label{eq:contup}
\end{equation}
and either for the rigid walls of case a);
\begin{equation}
 \left(\pderiv{h}{\zeta}-\gamma F'(\zeta)\pderiv{h}{\xi}\right)_{\zeta=0,1}=0,\label{eq:walls}
\end{equation}
or for the periodic conditions of case b);
\begin{equation}
 G|_{\zeta=0}=G|_{\zeta=1}\e^{i \alpha k_{3}},\qquad  \left.\pderiv{G}{\zeta}\right|_{\zeta=0}=\left.\pderiv{G}{\zeta}\right|_{\zeta=1}\e^{i \alpha k_{3}}.\label{eq:periodic}
\end{equation}
\label{eq:govtransformed}
\end{subequations}

We have now completed the formulation of the mathematical model, which we shall solve in the following section.

\section{Analytic Solution}\label{sec:solution}
We solve \eqref{eq:govtransformed} by applying a Fourier transform in the $\xi$ variable,
\begin{equation}
 H(\lambda,\eta,\zeta)=\int_{-\infty}^{\infty}h(\xi,\eta,\zeta)\e^{i\lambda\xi}d\xi,
\end{equation}
and separating the solution $H(\lambda,\eta,\zeta)=Y(\lambda,\eta)Z(\lambda,\zeta)$ with separation constant $\tilde{\chi}$. This results in governing equations
\begin{equation}
 Y''+\left((d M)^{2}-\lambda^{2}-\tilde{\chi}^{2}\right)Y=0,
\end{equation}
and
\begin{equation}
 Z''+2i\gamma\lambda F'(\zeta)Z'+\tilde{\chi}^{2}Z=-i\lambda\gamma\left(\delta(\zeta-\frac{3}{4})-\delta(\zeta-\frac{1}{4})\right)Z,
\label{eq:Zeq}
\end{equation}
for the $\eta$ and $\zeta$ dependencies. 

We solve \eqref{eq:Zeq} by considering an ansatz of the form
\begin{equation}
 Z(\lambda,\zeta)=\e^{-i\gamma\lambda F(\zeta)}\left(A(\lambda)\cos(\chi\zeta)+B(\lambda)\sin(\chi\zeta)\right),
\end{equation}
 which is found to satisfy \eqref{eq:Zeq} when $\chi^{2}=\lambda^{2}\gamma^{2}+\tilde{\chi}^{2}$. This yields solutions for $Y$ given by
\begin{equation}
 Y(\lambda,\eta)=\text{sgn}(\eta)\e^{-|\eta|\sqrt{1-\gamma^{2}}\sqrt{\lambda^{2}-w^{2}}},
\end{equation}
where 
\begin{equation}
 w^{2}=\frac{(d M)^{2}-\chi^{2}}{1-\gamma^{2}}.\label{eq:wn}
\end{equation}

To determine suitable values for $\chi$ and a relationship between $A(\lambda)$ and $B(\lambda)$ we must apply the boundary conditions to $Z$ at $\zeta=0,1$. These give rise to a modal expansion of the solution indexed by $n$.
Using the rigid walled condition, \eqref{eq:walls}, yields
\begin{equation}
 Z(\lambda,\zeta)=A_{n}(\lambda)Z_{n}(\lambda,\zeta)=\e^{-i\gamma\lambda F(\zeta)}A_{n}(\lambda)\cos(n\upi \zeta),\quad \chi=n\upi,\quad n\in\mathbb{Z},\label{eq:Zwalls}
\end{equation}
whilst the periodic condition, \eqref{eq:periodic}, yields
\begin{equation}
 Z(\lambda,\zeta)=A_{n}(\lambda)Z_{n}(\lambda,\zeta)=\e^{-i\gamma\lambda F(\zeta)}A_{n}(\lambda)\e^{-i k_{3}\alpha \zeta}\e^{2i n\upi \zeta},\quad \chi=\pm k_{3}\alpha+2n\upi,\quad n\in\mathbb{Z}.\label{eq:Zperiodic}
\end{equation}

We must now determine the $A_{n}(\lambda)$ using the $\eta=0$ boundary conditions and the Wiener-Hopf method.
We can write the general solution as
\begin{equation}
 H(\lambda,\eta,\zeta)=\sum_{n}A_{n}(\lambda)\text{sgn}(\eta)\e^{-|\eta|\sqrt{1-\gamma^{2}}\sqrt{\lambda^{2}-w_{n}^{2}}}Z_{n}(\lambda,\zeta),\label{eq:Hseries}
\end{equation}
where the $Z_{n}$ are either given in \eqref{eq:Zwalls} or \eqref{eq:Zperiodic}. The upstream continuity condition, \eqref{eq:contup}, tells us $A_{n}(\lambda)$ is a positive half-Fourier transform only, therefore is analytic in the upper half $\lambda-$plane which we denote by a superscript $+$ (analyticity in the lower half plane is similarly denoted by a superscript $-$). 

The zero normal velocity condition, \eqref{eq:zerovel}, upon applying the Fourier transform becomes
\begin{equation}
 \pderiv{H}{\eta}(\lambda,0,\zeta)=K^{+}(\lambda,\zeta)+U^{-}(\lambda,\zeta)
\end{equation}
where 
\begin{equation}
 K^{+}(\lambda,\zeta)=-\frac{i}{\lambda+\kappa}\e^{i\kappa\gamma F(\zeta)+i k_{3}\zeta}
\end{equation}
and $U^{-}(\lambda,\zeta)$ is an unknown function which is analytic in the lower half $\lambda-$plane.
Using \eqref{eq:Hseries} we obtain
\begin{equation}
 -\sum_{n}\sqrt{1-\gamma^{2}}\sqrt{\lambda^{2}-w_{n}^{2}}A_{n}^{+}(\lambda)Z_{n}(\lambda,\zeta)=-\frac{i}{\lambda+\kappa}\e^{i\kappa\gamma F(\zeta)+i k_{3}\zeta}+U^{-}(\lambda,\zeta).\label{eq:WHsum}
\end{equation}
The functions $Z_{n}(\lambda,\zeta)$ are orthogonal over the range $\zeta\in[0,1]$ with respect to their Schwartz conjugates thus we can use them as a basis for expanding our known and unknown functions.
In particular $U^{-}(\lambda,\zeta)$ can be expressed as
\begin{equation}
U^{-}(\lambda,\zeta)=\sum_{n}D_{n}(\lambda)Z_{n}(\lambda,\zeta)
\end{equation}
and we write $\e^{i\kappa\gamma F(\zeta)+i k_{3}\zeta}$ as
\begin{equation}
\e^{i\kappa\gamma F(\zeta)+i k_{3}\zeta}=\sum_{n}E_{n}(\lambda)Z_{n}(\lambda,\zeta).\label{eq:BCexpand}
\end{equation}

The functions $E_{n}(\lambda)$ arise because of the spanwise form of the normal velocity on the plate and are given in the Appendix \ref{app:1}.
If we suppose, like \citet{Envia}, that the normal velocity just upstream of the plate must have a similar spanwise $\zeta$ dependence, then by linearity each $A_{n}^{+}$ and each $D_{n}^{-}$ must contain a factor of $E_{n}(\lambda)$ \footnote{To assume this we must also neglect any boundary layer effects due to the channel walls at $\zeta=0,1$. We believe this is a suitable assumption as we are interested not in the effects of the channel walls, but solely in the effects of the serration}.

We factor out $E_{n}$ in our Wiener-Hopf equation to obtain

\begin{equation}
\sqrt{1-\gamma^{2}}\sqrt{\lambda^{2}-w_{n}^{2}}\tilde{A}_{n}^{+}(\lambda)=\frac{i}{\lambda+\kappa}+\tilde{D}_{n}(\lambda),\label{eq:WHexp}
\end{equation}
where 
\begin{equation}
\tilde{A}_{n}^{+}E_{n}=A_{n}^{+},\qquad \tilde{D}_{n}^{-}E_{n}=D_{n}^{-}
\end{equation}

The $E_{n}(\lambda)$ are entire, therefore we can factor them out of the terms $A^{+}_{n}$ and $D^{-}_{n}$ without affecting the domain of analyticity.

We rearrange \eqref{eq:WHexp} to give
\begin{equation}
\sqrt{1-\gamma^{2}}\sqrt{\lambda+w_{n}}\tilde{A}_{n}^{+}(\lambda)=\frac{i}{\lambda+\kappa}\frac{1}{\sqrt{\lambda-w_{n}}}-\frac{D_{n}^{-}(\lambda)}{\sqrt{\lambda-w_{n}}}.
\end{equation}
The left hand side is analytic in the upper half $\lambda-$plane, and the unknown term on the right hand side is analytic in the lower half $\lambda-$plane. By additively splitting the known term on the right hand side into two functions, $F^{+}_{n}+F^{-}_{n}$, that are analytic in the appropriate half planes we can apply Liouville's theorem to solve for $\tilde{A}^{+}_{n}(\lambda)$ giving
\begin{equation}
\tilde{A}_{n}^{+}(\lambda)=\frac{F^{+}_{n}(\lambda)}{\sqrt{1-\gamma^{2}}\sqrt{\lambda+w_{n}}},
\end{equation}
with 
\begin{equation}
 F^{+}_{n}(\lambda)=\frac{i}{\lambda+\kappa}\frac{1}{\sqrt{-\kappa-w_{n}}},
\label{eq:fplus}
\end{equation}
hence 
\begin{equation}
H(\lambda,\eta,\zeta)=\text{sgn}(\eta)\sum_{n}\frac{F^{+}_{n}(\lambda)E_{n}(\lambda)\e^{-|\eta|\sqrt{1-\gamma^{2}}\sqrt{\lambda^{2}-w_{n}^{2}}}}{\sqrt{1-\gamma^{2}}\sqrt{\lambda+w_{n}}}Z_{n}(\lambda,\zeta).
\end{equation}

We invert the Fourier transform and obtain the far-field ($r\gg1$) acoustics by applying the method of steepest descents to give
\begin{align}
&h(r,\theta,z)\sim\notag
\\
&\sum_{n}\frac{\e^{\upi i/4}F^{+}_{n}(-w_{n}\cos\theta)E_{n}(-w_{n}\cos\theta)}{(1-\gamma^{2})^{3/4}\sqrt{\upi}}\cos\left(\frac{\theta}{2}\right)\frac{\e^{i\sqrt{1-\gamma^{2}}w_{n}r}}{\sqrt{r}}Z_{n}(-w_{n}\cos\theta,z)\e^{-i\gamma w_{n}\cos\theta F(z)},\label{eq:hff}
\end{align}
where $(r,\theta,z)$ are standard cylindrical polar coordinates with origin corresponding to Cartesian origin $x=y=z=0$. For simplicity, we write \eqref{eq:hff} as
\begin{equation}
h(r,\theta,z)\sim\sum_{n}B_{n}(\theta,z)\frac{\e^{i\sqrt{1-\gamma^{2}}w_{n}r}}{\sqrt{r}},\label{eq:hff2}
\end{equation}
where we refer to the $B_{n}$ as the scattered modes with frequencies $\sqrt{1-\gamma^{2}}w_n$.
We recall the definition of $w_{n}$ in \eqref{eq:wn} and see that for sufficiently large $n$ these modes are cutoff, therefore practically we only need to sum a finite number of terms in \eqref{eq:hff2} to calculate this far-field expression.

To obtain the acoustic pressure, we use the relation
\begin{equation}
p=-\left(\pderiv{h}{x}-\frac{i k_{1}}{\beta^{2}}h\right)\e^{-i k_{1}M^{2}x/\beta^{2}},
\end{equation}
which in the far field, $r\gg1$, using \eqref{eq:hff2} yields
\begin{equation}
p(r,\theta,z)\sim \sum_{n}i\left(\frac{k_{1}}{\beta^{2}}-\sqrt{1-\gamma^{2}}w_{n}\right)B_{n}(\theta,z)\frac{\e^{i\sqrt{1-\gamma^{2}}w_{n}r}}{\sqrt{r}}.\label{eq:pff}
\end{equation}

This completes our analytic solution of the scattered field.

\section{Results}\label{sec:results}

\begin{figure}
\centering
 \includegraphics[clip, trim= 0cm 5cm 0cm 5cm, width=0.65\textwidth]{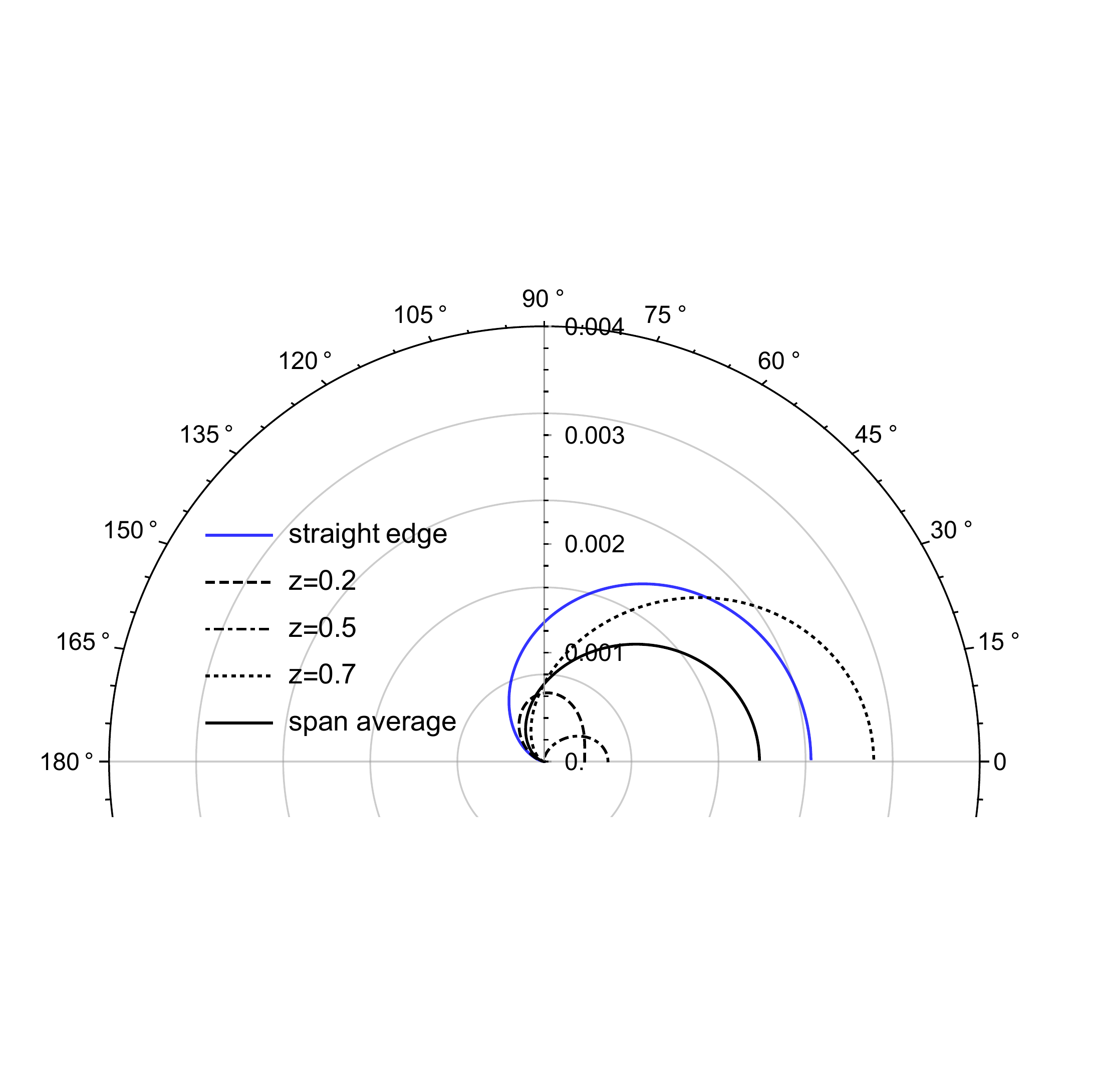}
\caption{Polar plot of $|p(10,\theta,z)|^2$ compared to the span average, $D_a(10,\theta)$, and straight-edge,$c=0$, results, for blades in a rigid walled channel. In all cases $M=0.3$, $k_{3}=0$, $k_{1}=10$. Serrated blades have $c=1$.}
\label{fig:spanvary}
\end{figure}

In this section we present far-field noise results for both the rigid-walled channel and periodic channel.
To do so, we define a spanwise averaged far-field pressure, as 
\begin{equation}
D_{a}(r,\theta)=\int_{0}^{1}|p(r,\theta,z)|^{2}dz,
\label{eq:avedirect}
\end{equation}
where $p(r,\theta,z)$ is given by \eqref{eq:pff}. We evaluate this numerically, using Mathematica's inbuilt NIntegrate feature. We only sum a finite number of terms from $\eqref{eq:pff}$ corresponding to the cuton terms with $w_{n}\in\mathbb{R}$. The number of cuton terms depends on the gust wavenumber components, $k_{1,3}$, and the Mach number of the background flow, $M$. Figure \ref{fig:spanvary} illustrates a variation in $|p(r,\theta,z)|^2$ with spanwise location, $z$. Different spanwise locations can produce different directivity patterns at different magnitudes due to interference between different $B_{n}$ modes. We see in Figure \ref{fig:spanvary} that the spanwise averaged pressure recovers the cardiod pattern typically associated with a straight-edge interaction, and we believe the spanwise average gives the best indication of the overall noise generated in the channel.

The layout of this results section is as follows. First in section \ref{sec:rigid} we replicate the results from \citet{Envia} for a swept blade in a rigid walled channel and compare to our serrated blade in a channel. Second in section \ref{sec:periodic} we present results for the far-field noise from a serrated edge in a periodic channel, and we discuss how the far-field results are affected by the channel wall conditions, with attention given to noise generated by gusts with both zero and non-zero spanwise wavenumbers, $k_{3}$. By noting the importance of the spanwise wavenumber on the far-field sound, in section \ref{sec:PSD} we integrate over a spectrum of $k_{3}$ values to compare the far-field power spectral density calculated analytically to that measured experimentally from \citep{sn15}. We use the analytic results to infer noise reduction mechanisms for the serrated leading edge in section \ref{sec:reds}. Finally in section \ref{sec:numerics} we compare the analytic results against numerical predictions.

\subsection{Spanwise averaged far-field pressure in a rigid walled channel}\label{sec:rigid}

\begin{figure}
 \centering
\begin{subfigure}[b]{0.49\linewidth}
 \centering
 \includegraphics[width=1.1\textwidth]{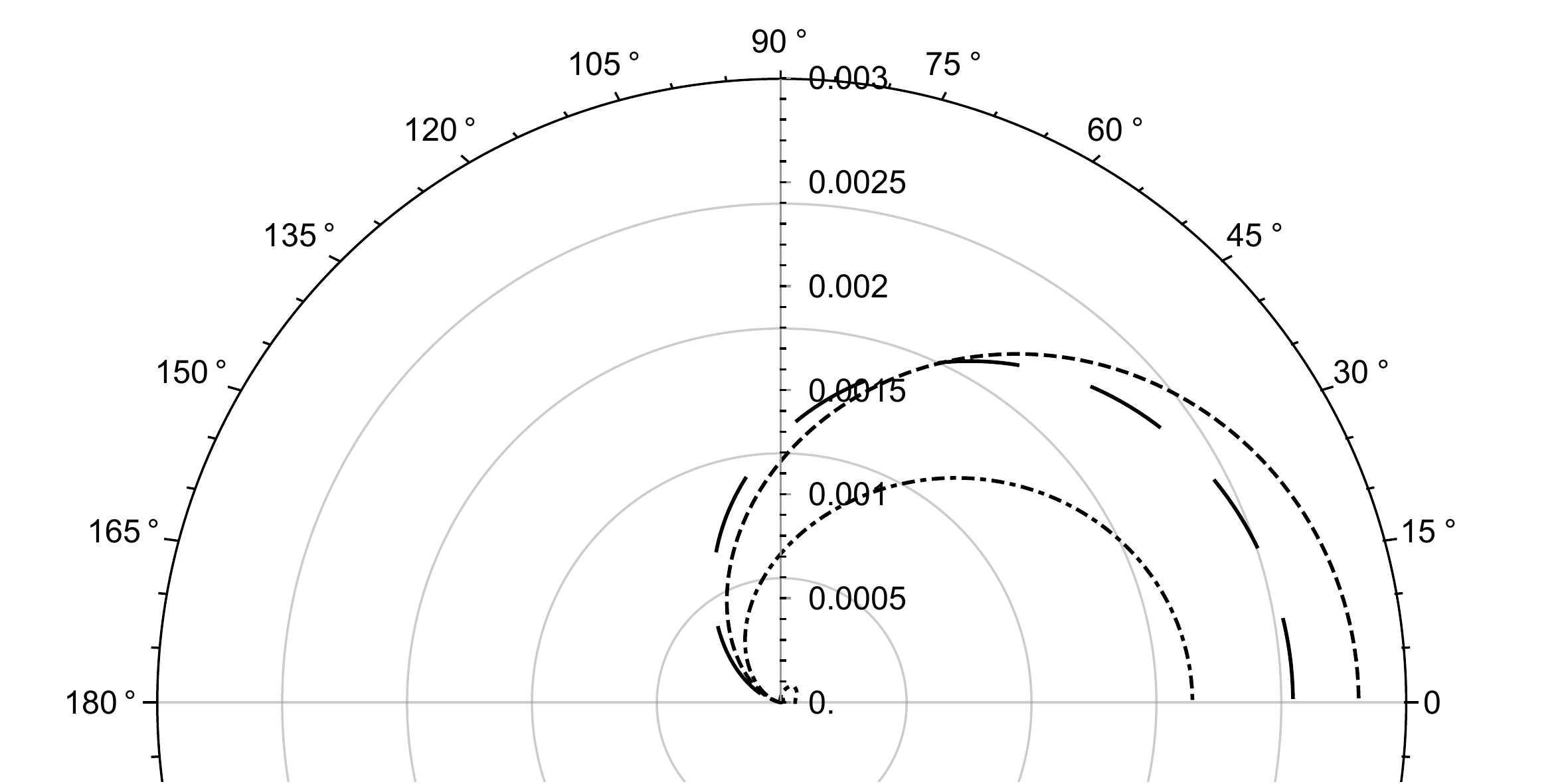}
\caption{$k_{1}=10$, Case a).}
\end{subfigure}
\hfill
\begin{subfigure}[b]{0.49\linewidth}
 \centering
 \includegraphics[width=1.2\textwidth]{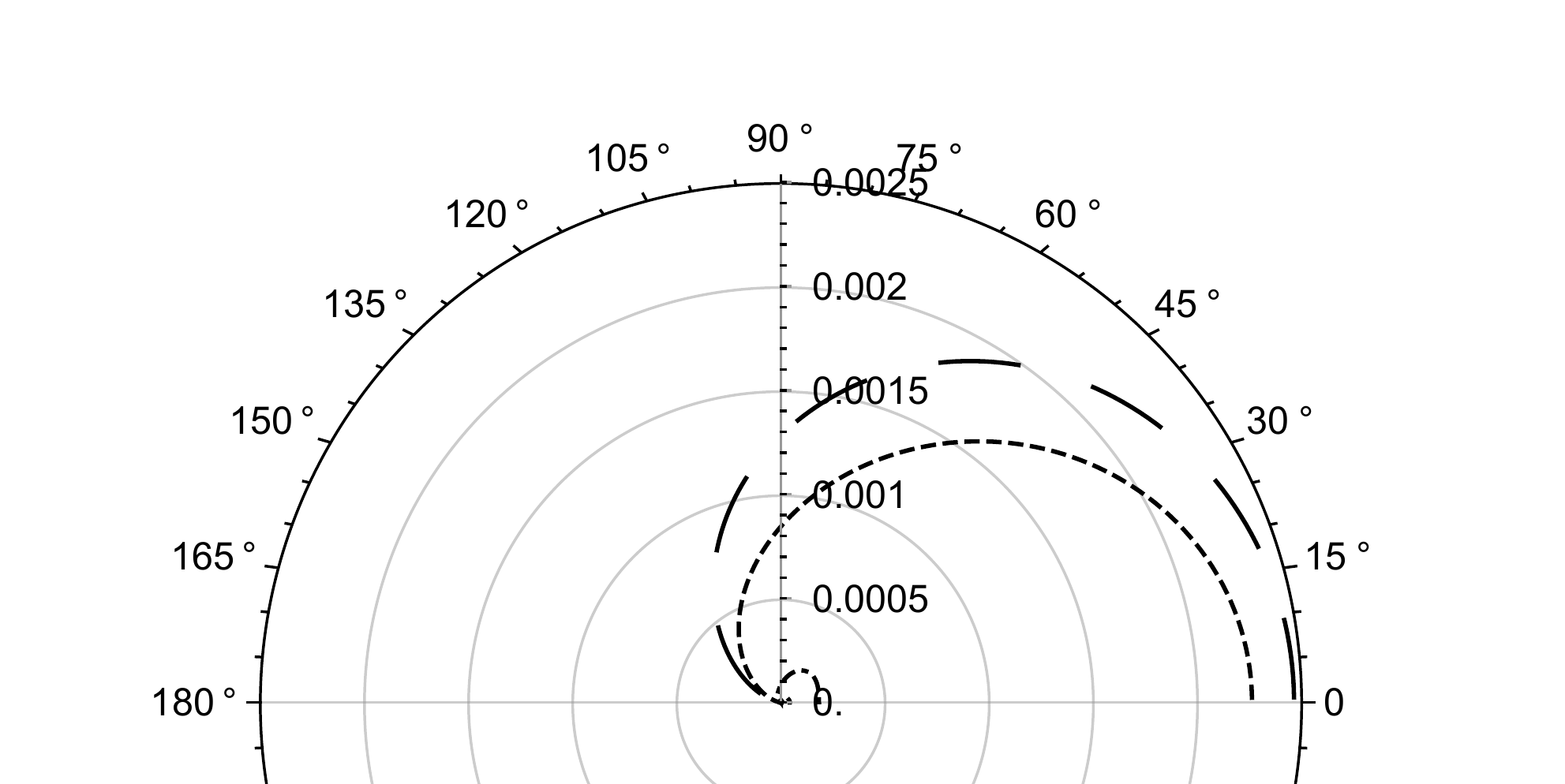}
\caption{$k_{1}=10$, swept blade.}
\end{subfigure}
\\
\vspace{0.05\linewidth}
\begin{subfigure}[b]{0.49\linewidth}
 \centering
\includegraphics[width=1.1\textwidth]{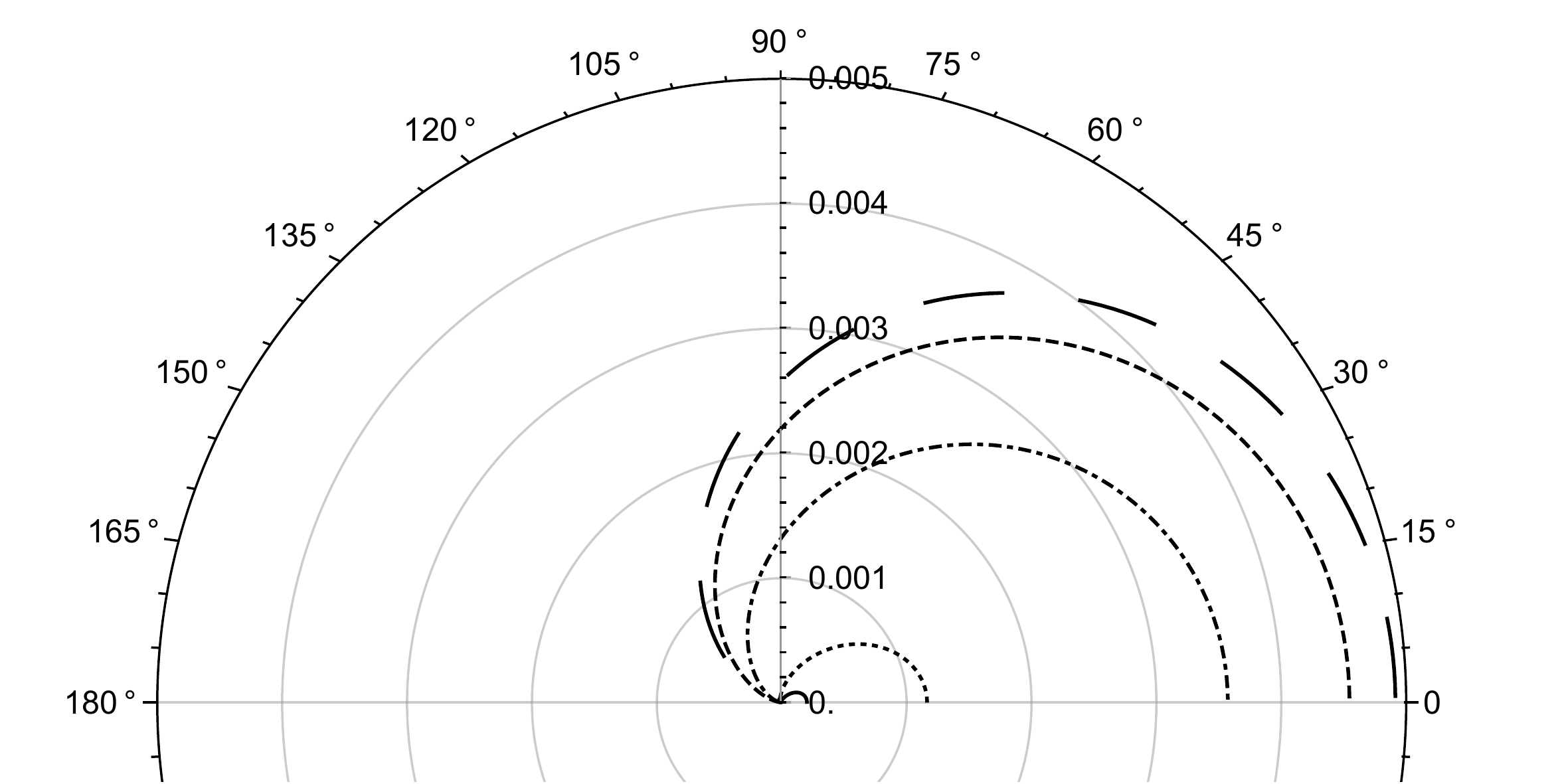}
\caption{$k_{1}=5$, Case a).}
\end{subfigure}
\hfill
\begin{subfigure}[b]{0.49\linewidth}
 \centering
 \includegraphics[width=1.1\textwidth]{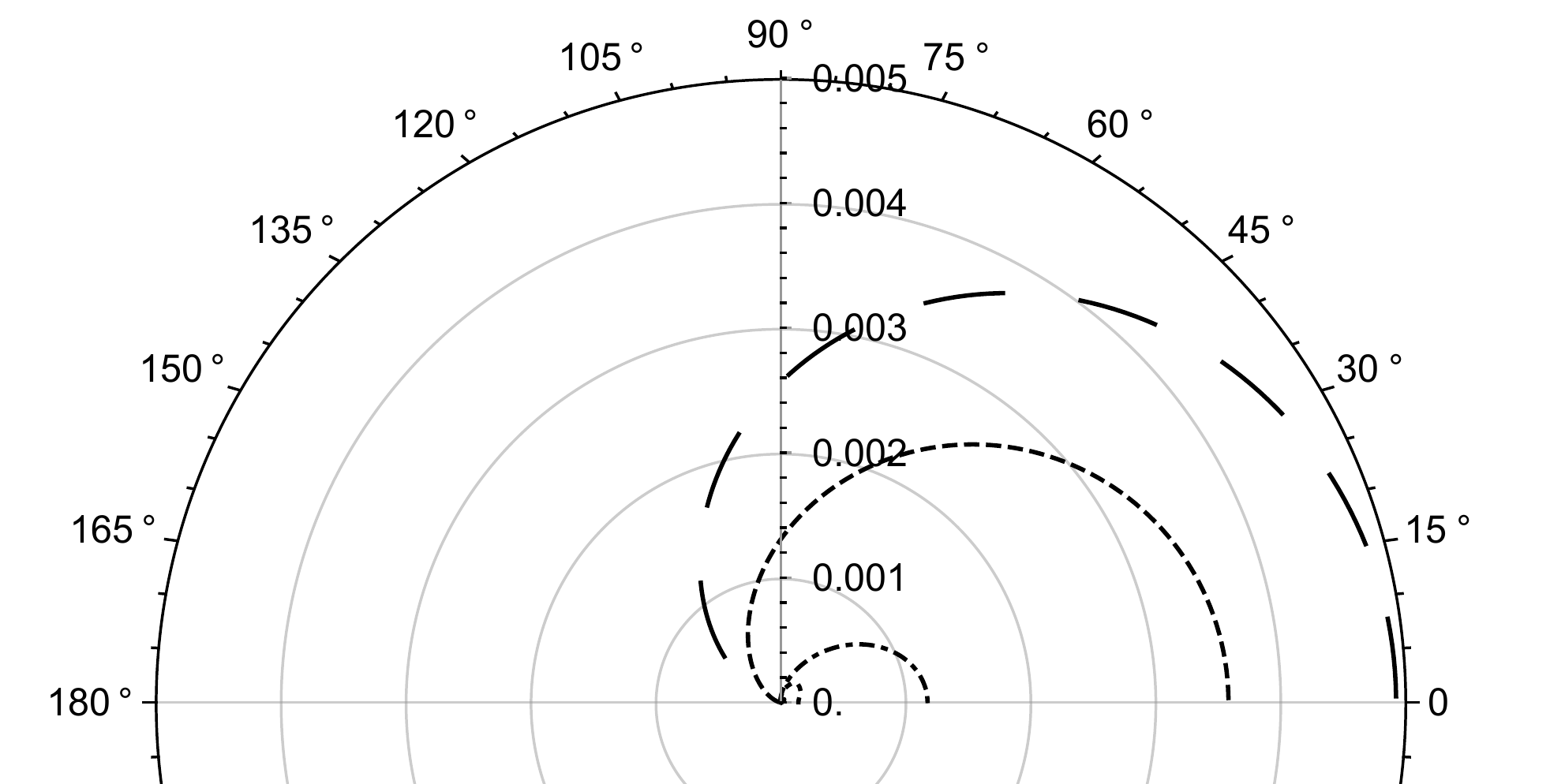}
\caption{$k_{1}=5$, swept blade.}
\end{subfigure}
\\
\vspace{0.05\linewidth}
\begin{subfigure}[b]{0.49\linewidth}
 \centering
 \includegraphics[width=1.1\textwidth]{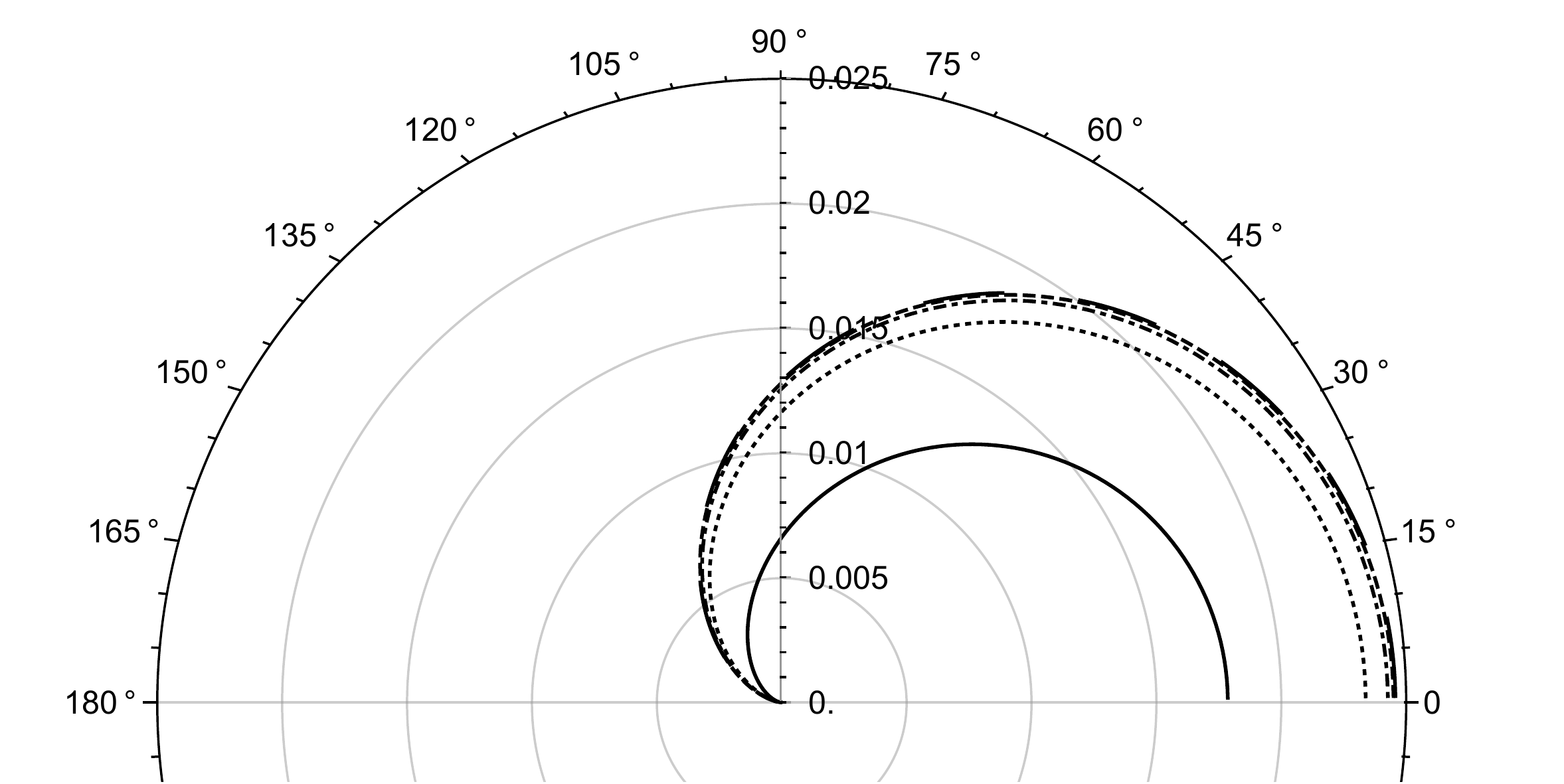}
\caption{$k_{1}=1$, Case a).}
\end{subfigure}
\hfill
\begin{subfigure}[b]{0.49\linewidth}
 \centering
  \includegraphics[width=1.1\textwidth]{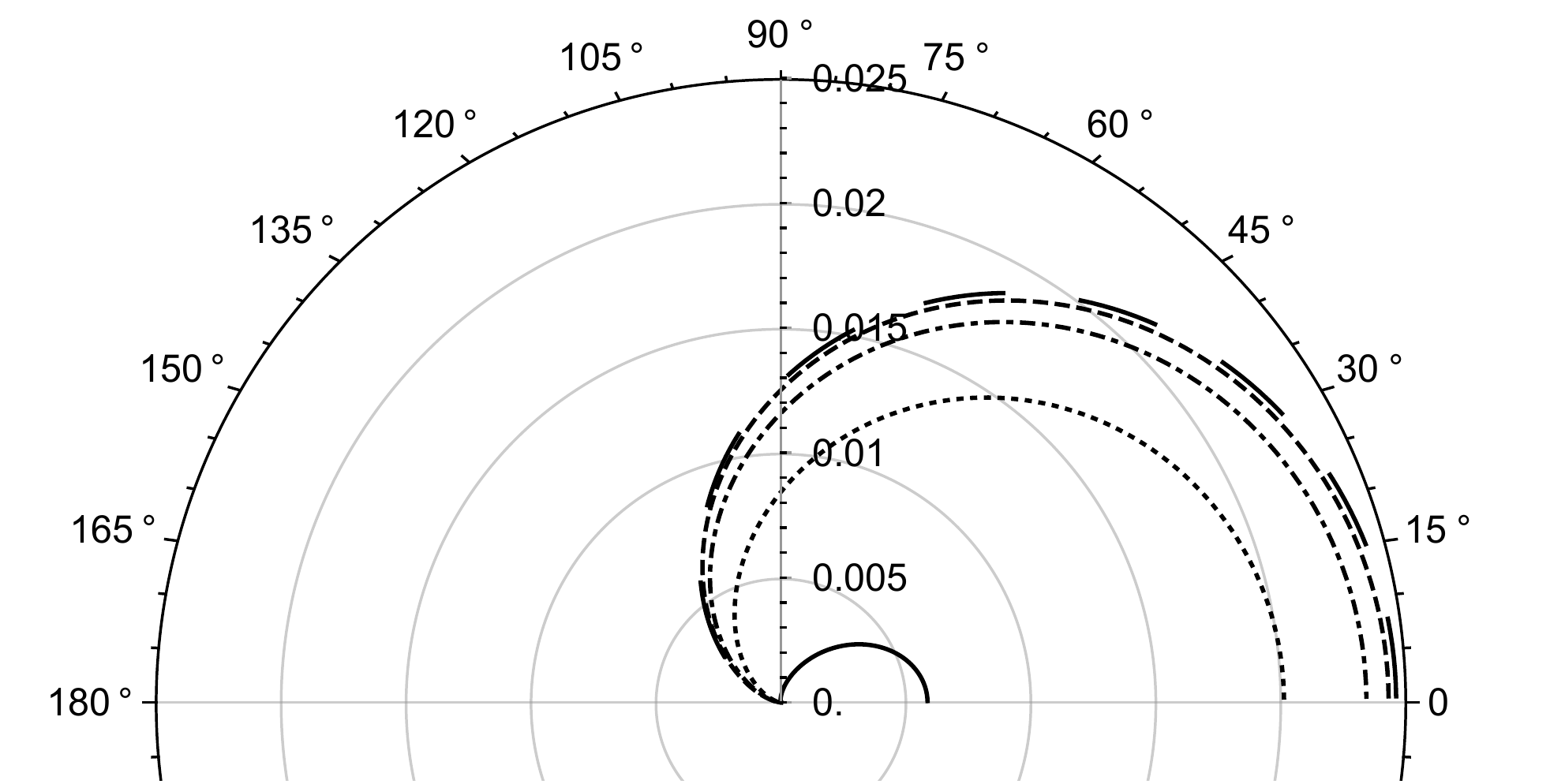}
\caption{$k_{1}=1$, swept blade.}
\end{subfigure}
\\
\vspace{0.05\linewidth}
\begin{subfigure}[b]{0.49\linewidth}
 \centering
 \includegraphics[width=1.1\textwidth]{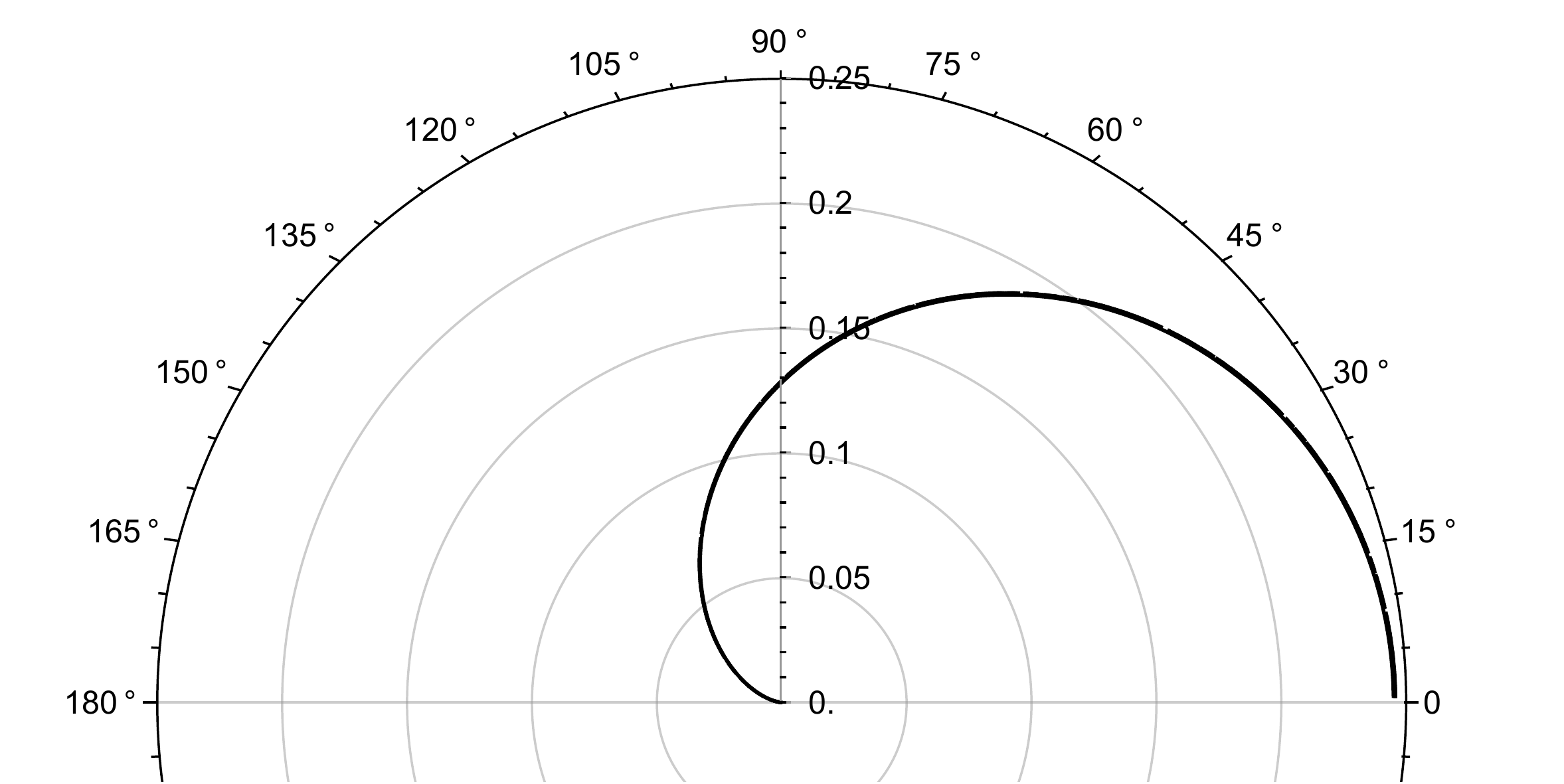}
\caption{$k_{1}=0.1$, Case a).}
\end{subfigure}
\hfill
\begin{subfigure}[b]{0.49\linewidth}
 \centering
 \includegraphics[width=1.1\textwidth]{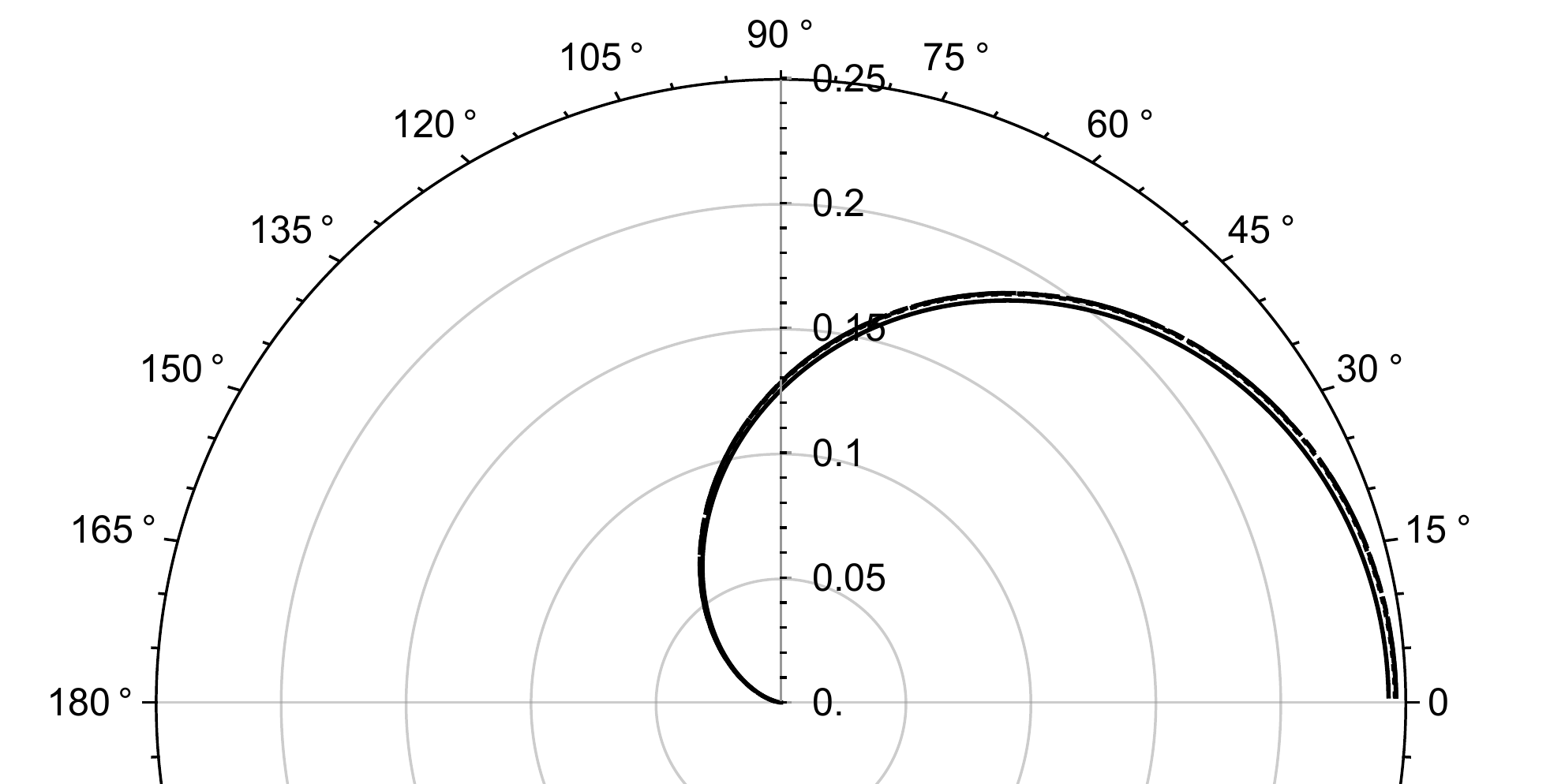}
\caption{$k_{1}=0.1$, swept blade.}
\end{subfigure}
\caption{Polar plot of the spanwise average directivity, $D_{a}(r,\theta)$ as given by \eqref{eq:avedirect}, for $r=10$, $M=0.3$, $k_{3}=0$. Large dashed $c=0$; dashed $c=0.5$; dot-dashed $c=1$; dotted $c=2$, solid $c=5$.}
\label{fig:sweepcompare}
\end{figure}

Here we compare the span averaged directivity for a serrated blade in a channel to a swept blade in a channel (known from \citet{Envia}) in Figure \ref{fig:sweepcompare}. We see that the swept blade is more effective at reducing the far-field noise than the serration across all frequency ranges as the gradient of the edges (sweep angle or serration tip-to-root height) increase. To understand this we consider the individual modes contributing to the solutions in each case. 

\begin{figure}
 \centering
\begin{subfigure}[b]{0.49\linewidth}
 \centering
 \includegraphics[width=1.1\textwidth]{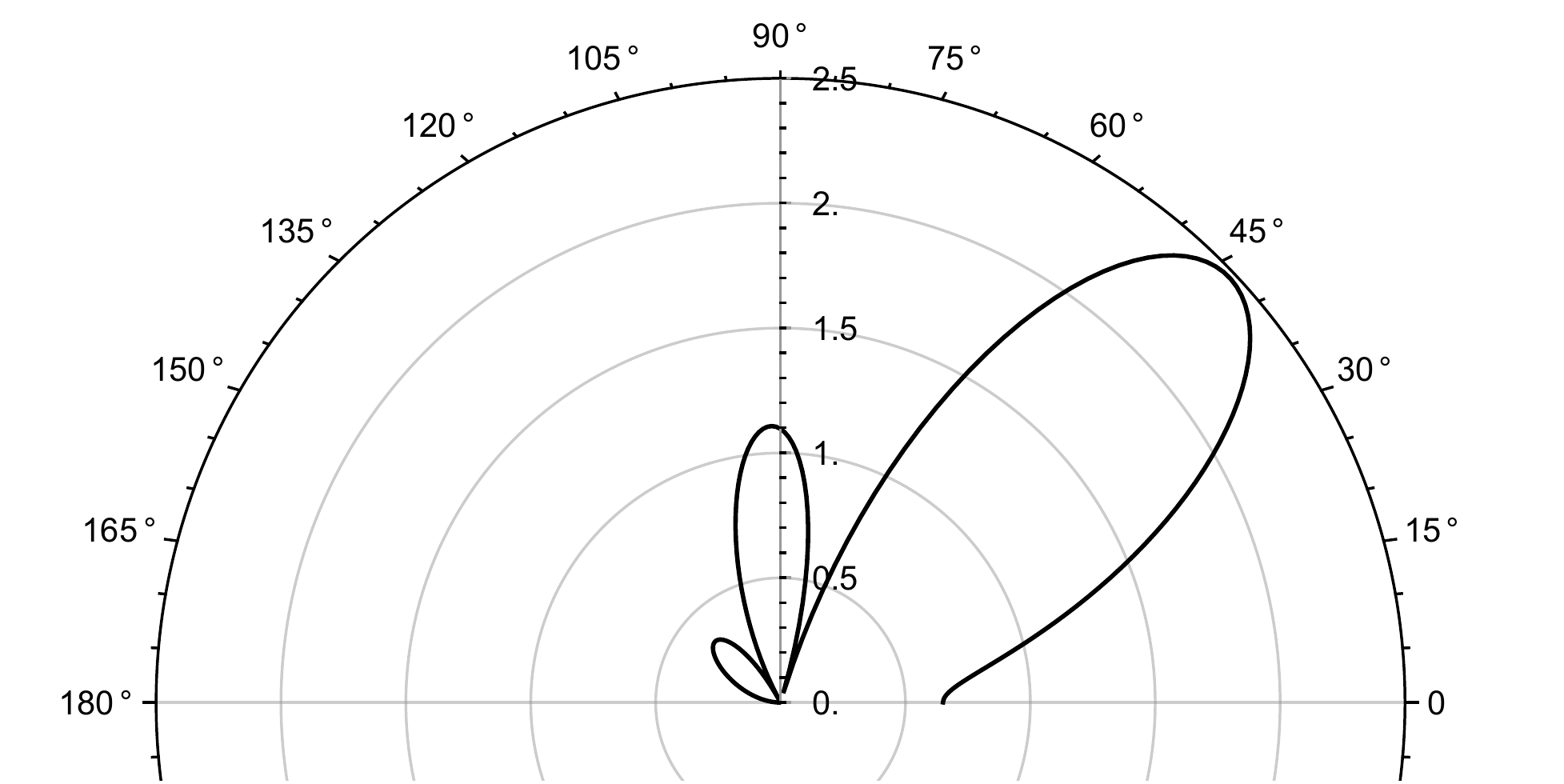}
\caption{$|B_{0}(\theta,0.5)|$, Case a).}
\label{fig:serrationmode0}
\end{subfigure}
\hfill
\begin{subfigure}[b]{0.49\linewidth}
 \centering
 \includegraphics[width=1.1\textwidth]{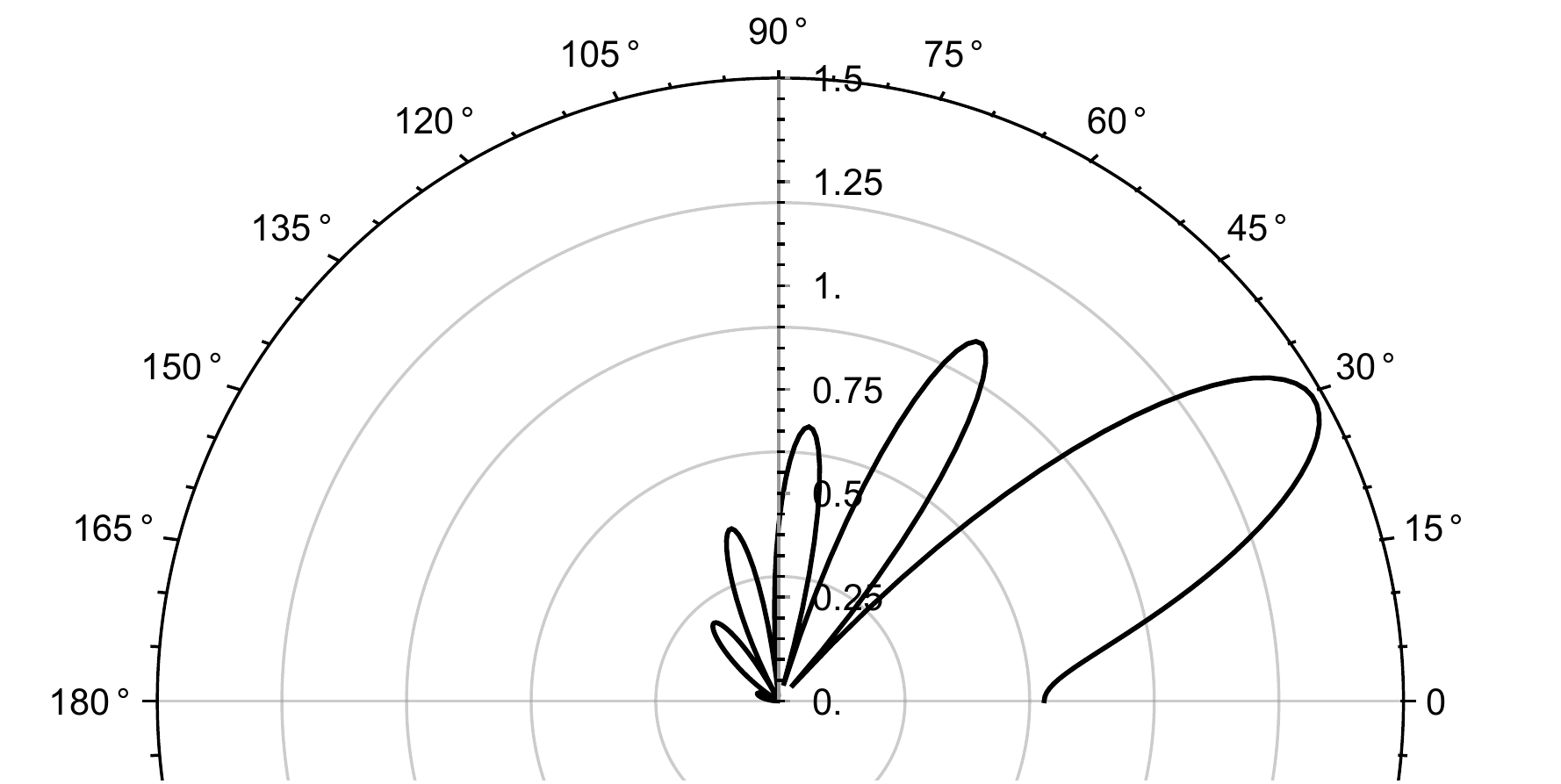}
\caption{$|B_{0}(\theta,0.5)|$, swept blade.}
\end{subfigure}
\\
\vspace{0.05\linewidth}
\begin{subfigure}[b]{0.49\linewidth}
 \centering
\includegraphics[width=1.1\textwidth]{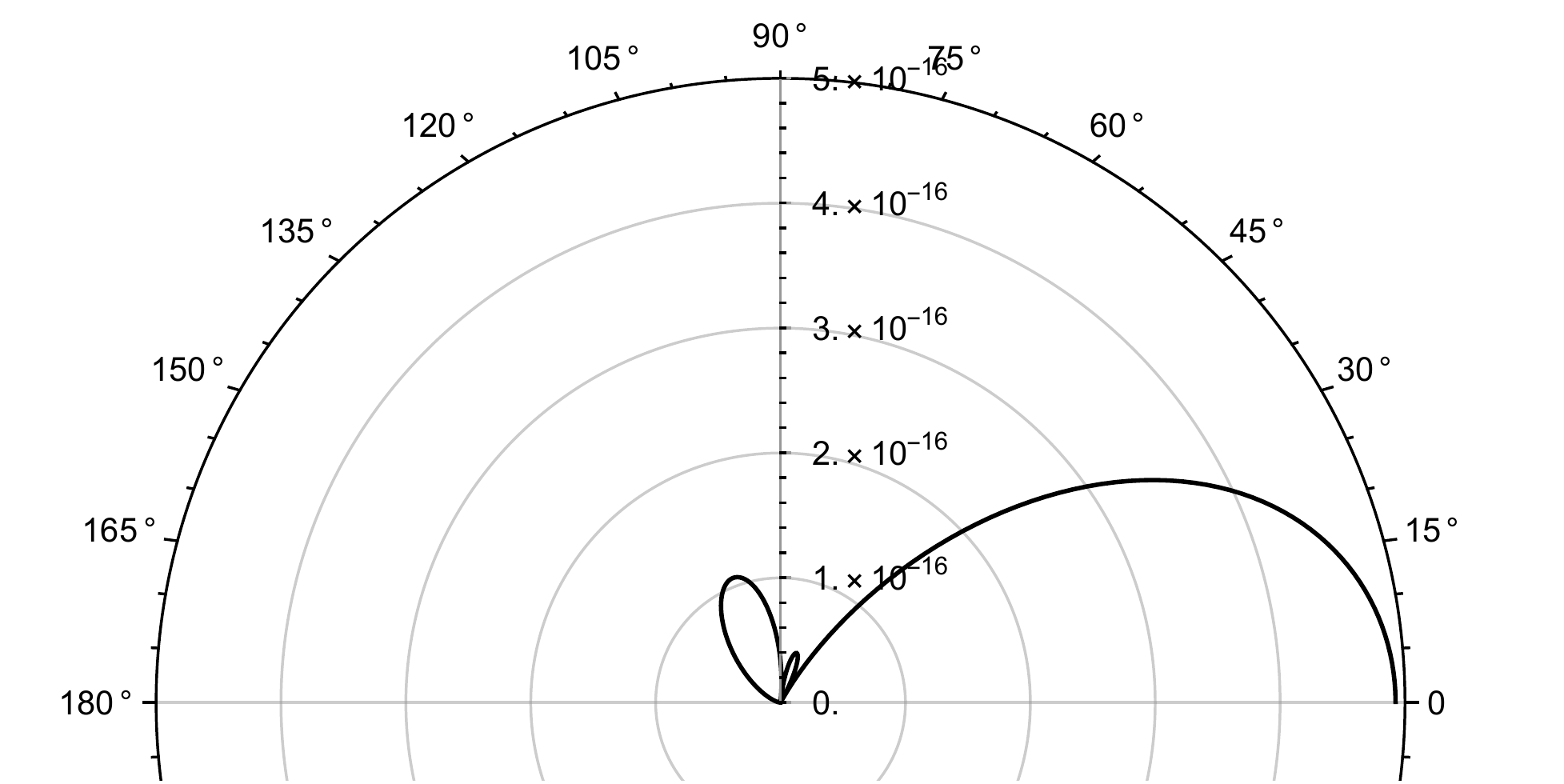}
\caption{$|B_{2}(\theta,0.5)|$, Case a).}
\end{subfigure}
\hfill
\begin{subfigure}[b]{0.49\linewidth}
 \centering
 \includegraphics[width=1.1\textwidth]{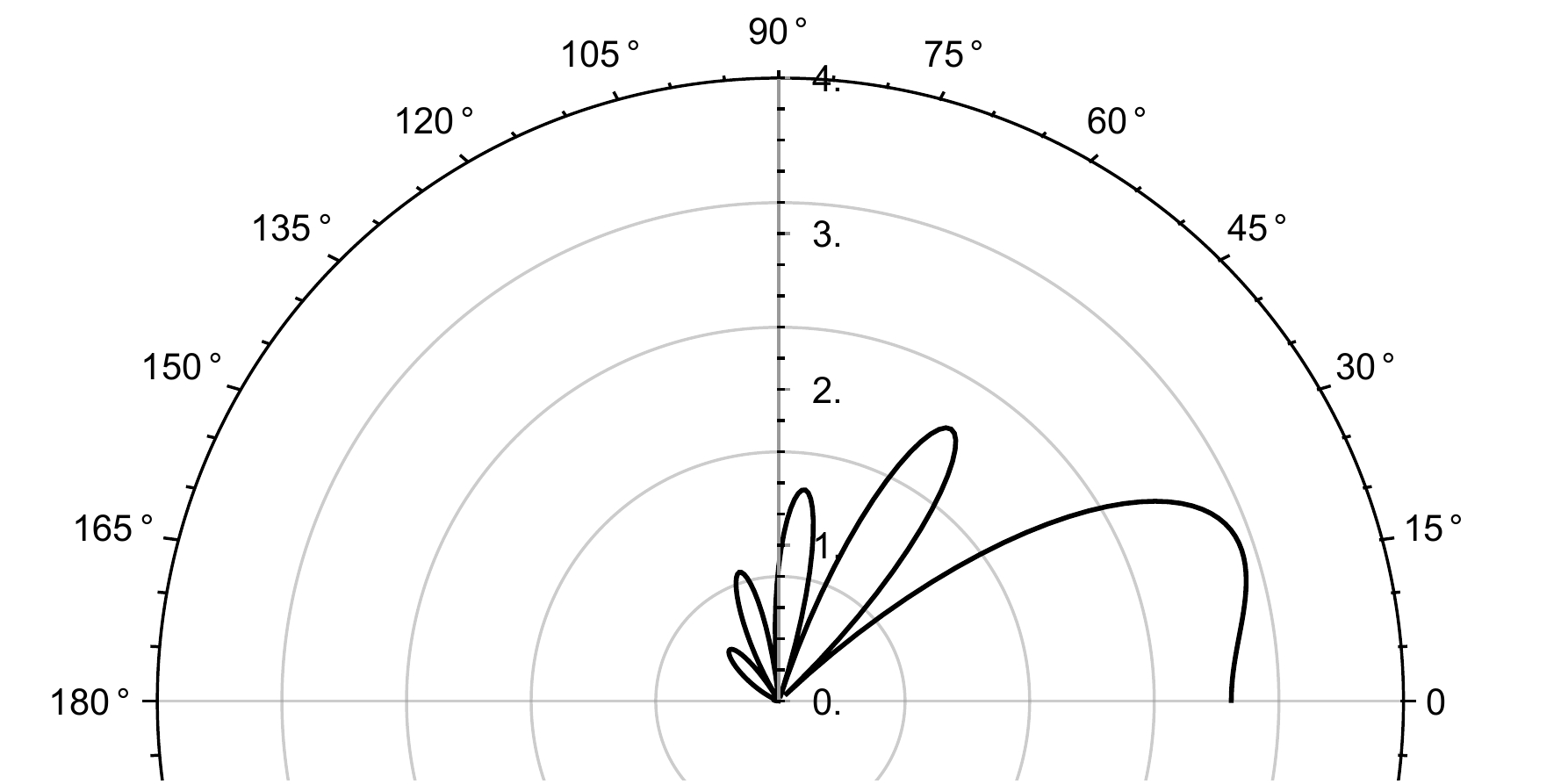}
\caption{$|B_{2}(\theta,0.5)|$, swept blade.}
\end{subfigure}
\\
\vspace{0.05\linewidth}
\begin{subfigure}[b]{0.49\linewidth}
 \centering
 \includegraphics[width=1.1\textwidth]{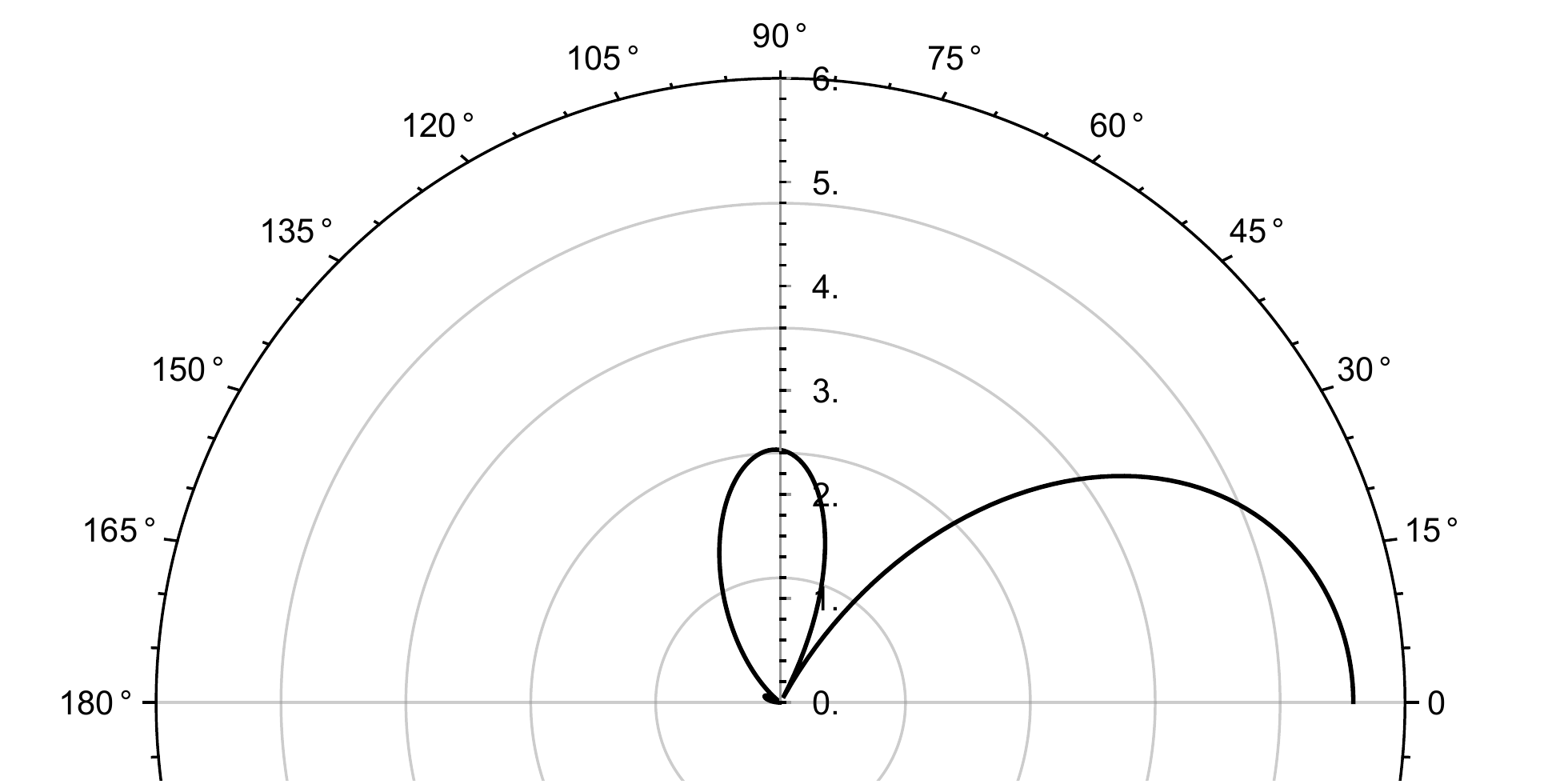}
\caption{$|B_{4}(\theta,0.5)|$, Case a).}
\end{subfigure}
\hfill
\begin{subfigure}[b]{0.49\linewidth}
 \centering
 \includegraphics[width=1.1\textwidth]{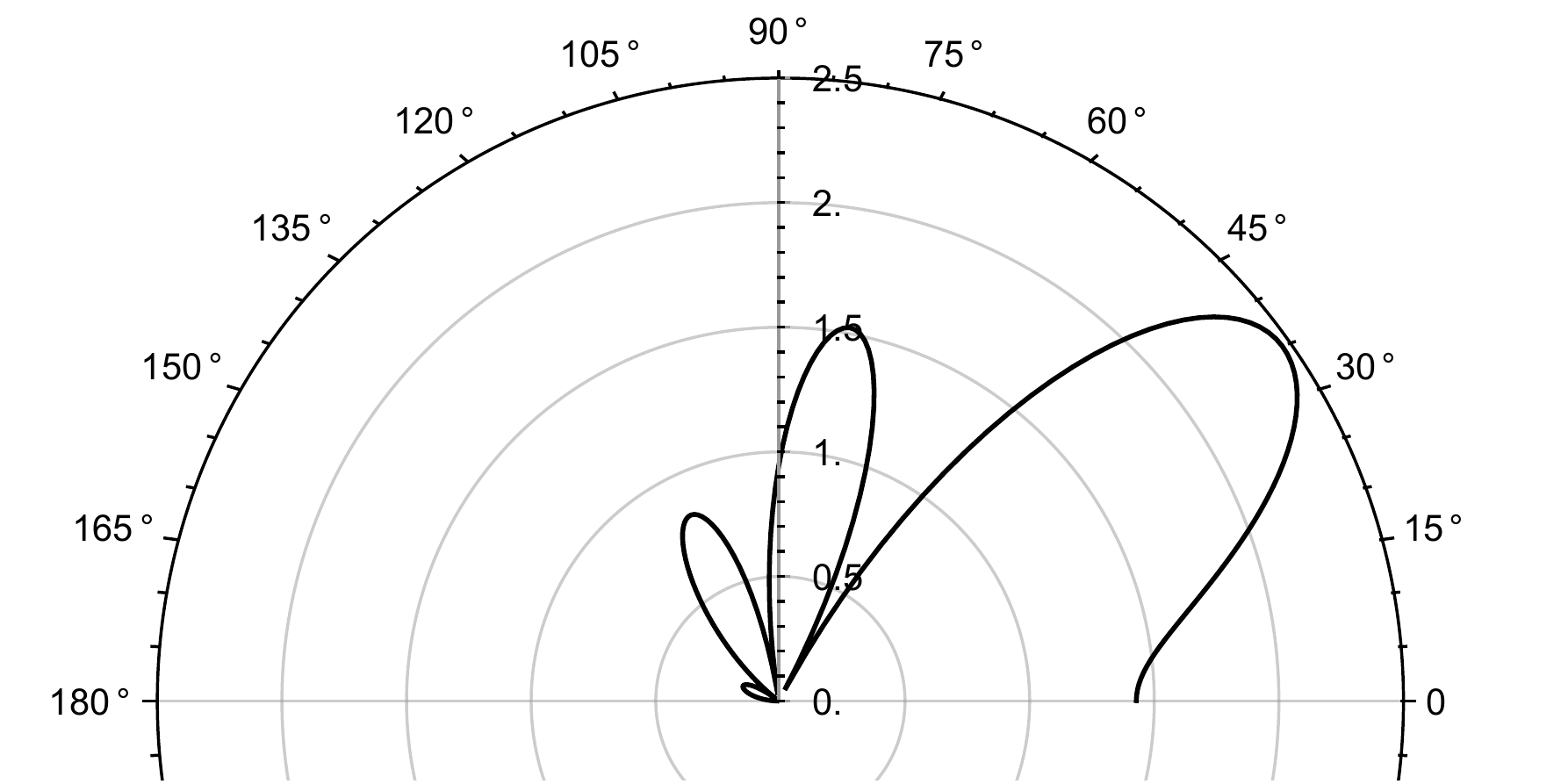}
\caption{$|B_{4}(\theta,0.5)|$, swept blade.}
\end{subfigure}
\caption{Polar plot of the magnitude of the $n$th mode, $10^4|B_{n}(\theta,z)|$, as defined in \eqref{eq:pff}, at mid-span point $z=0.5$, for $M=0.3$, $k_{3}=0$, $k_{1}=50$.}
\label{fig:modes}
\end{figure}

\begin{figure}
\centering
 \includegraphics[width=0.55\textwidth]{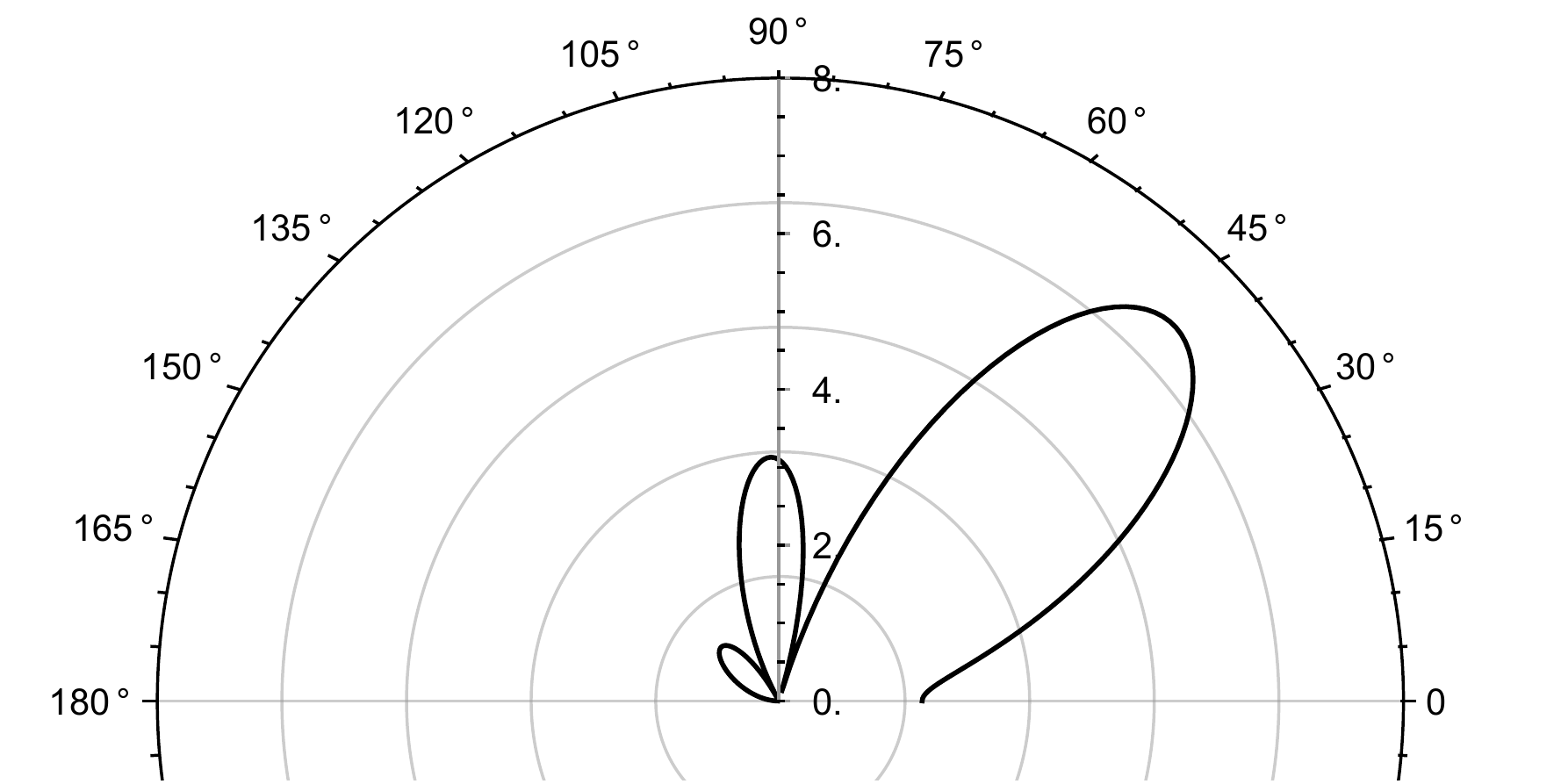}
\caption{Polar plot of $10^4|B_{0}(\theta,0.5)|$ for $M=0.3$, $k_{3}=0$, $k_{1}=25$ for a swept blade in a rigid channel.}
\label{fig:sweephalf}
\end{figure}

In Figure \ref{fig:modes} we plot the amplitude of the modes, $|B_{n}(\theta,z)|$, from \eqref{eq:pff}, contributing to the scattered field for $k_{1}=50$, $c=1$. We choose a very high frequency as this permits more scattered modes. We clearly see the modes for the swept blade are more oscillatory, thus encounter a greater destructive interference than the modes for the serrated edge. This is to be expected as the greatest horizontal distance between points for the swept blade is twice that of the serrated edge (the swept edge effectively has $F(z)=z$ rather than our definition of \eqref{eq:F}), and indeed the effective oscillations seen for the swept blade are twice those seen for the serrated edge (compare Figure \ref{fig:sweephalf} to Figure \ref{fig:serrationmode0}). This equivalence between the $0$th modes does not occur for higher modes, as these account for more complicated leading-edge geometry effects, and none of the modes are equivalent if $k_{3}\neq0$ as the modal expansions are heavily dependent on $k_{3}$ (spanwise variations in the flow expanded in different spanwise bases certainly shouldn't be equivalent). We also note that the oscillations in the modes decreases with increasing mode number; the zeroth mode is the most oscillatory. This is to be expected given that for mid-frequency interactions (such as $k_{1}=5$) only the zeroth mode propagates to the far field yet we see a highly oscillatory directivity for large $c$.

\subsection{Spanwise averaged far-field pressure in a periodic channel}\label{sec:periodic}

\begin{figure}
\centering
\begin{subfigure}[b]{0.49\linewidth}
 \centering
 \includegraphics[width=1.2\textwidth]{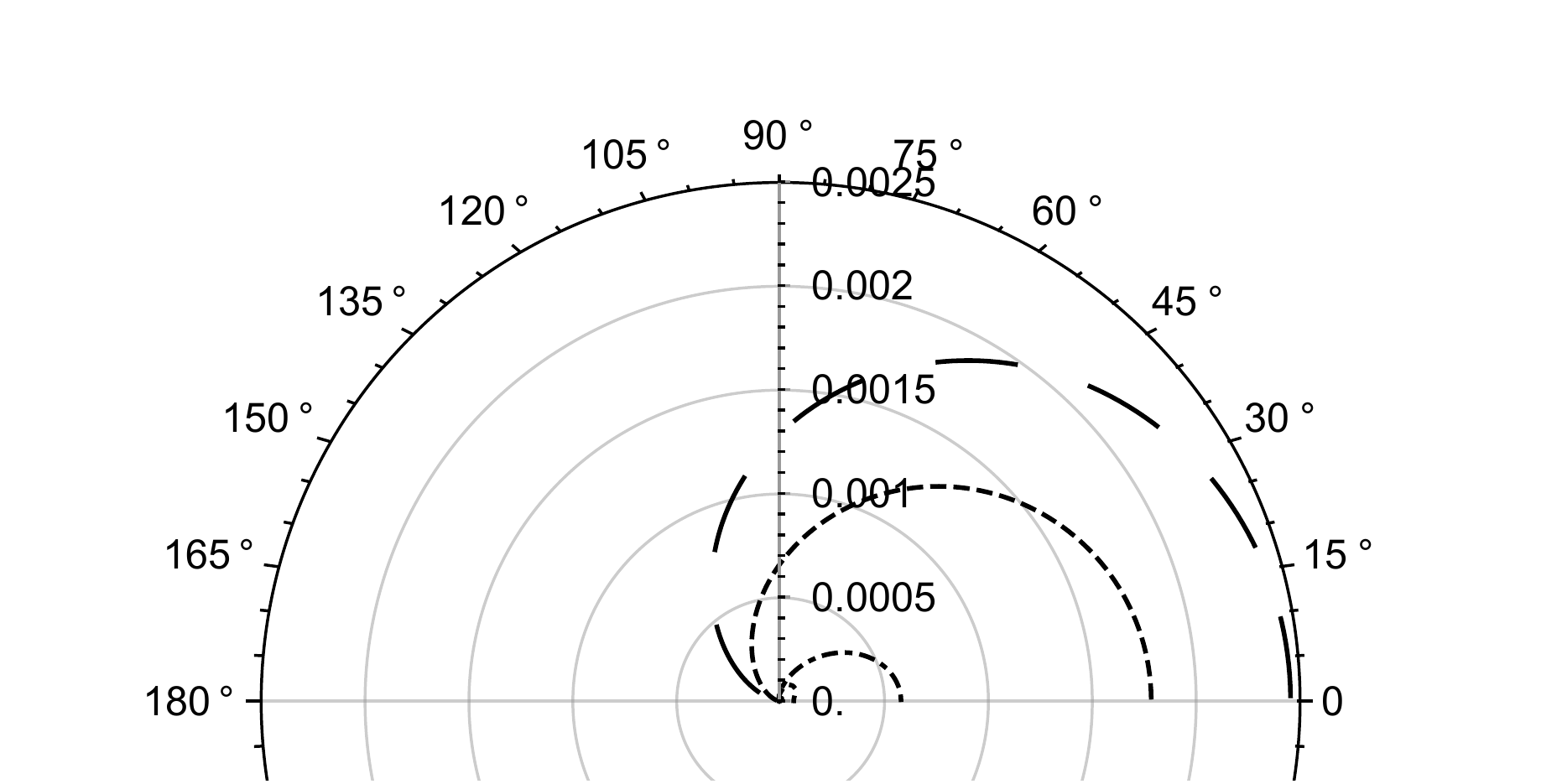}
\caption{$k_{1}=10$.}
\end{subfigure}
\hfill
\begin{subfigure}[b]{0.49\linewidth}
 \centering
 \includegraphics[width=1.1\textwidth]{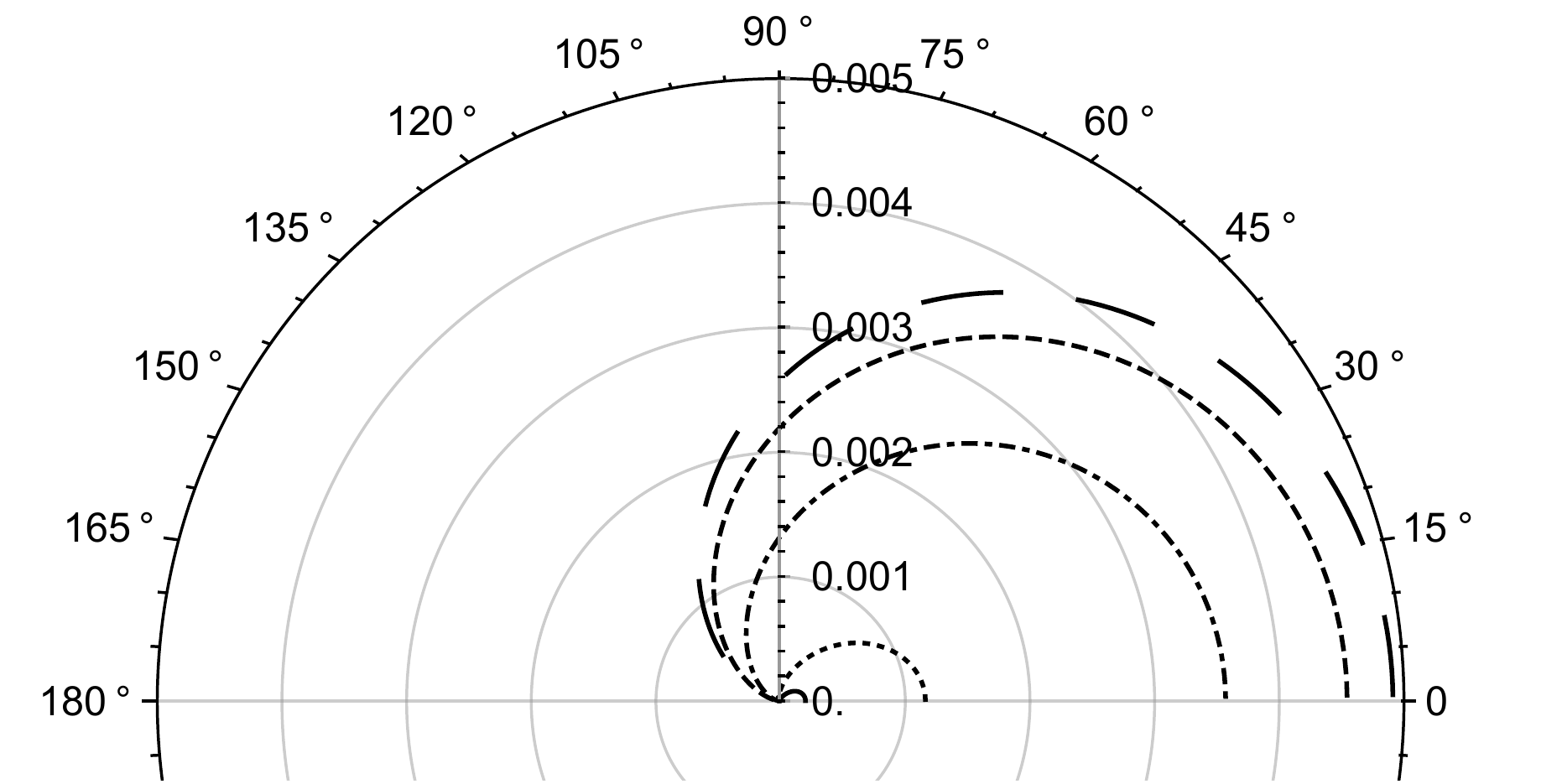}
\caption{$k_{1}=5$.}
\end{subfigure}
\\
\vspace{0.05\linewidth}
\begin{subfigure}[b]{0.49\linewidth}
 \centering
\includegraphics[width=1.1\textwidth]{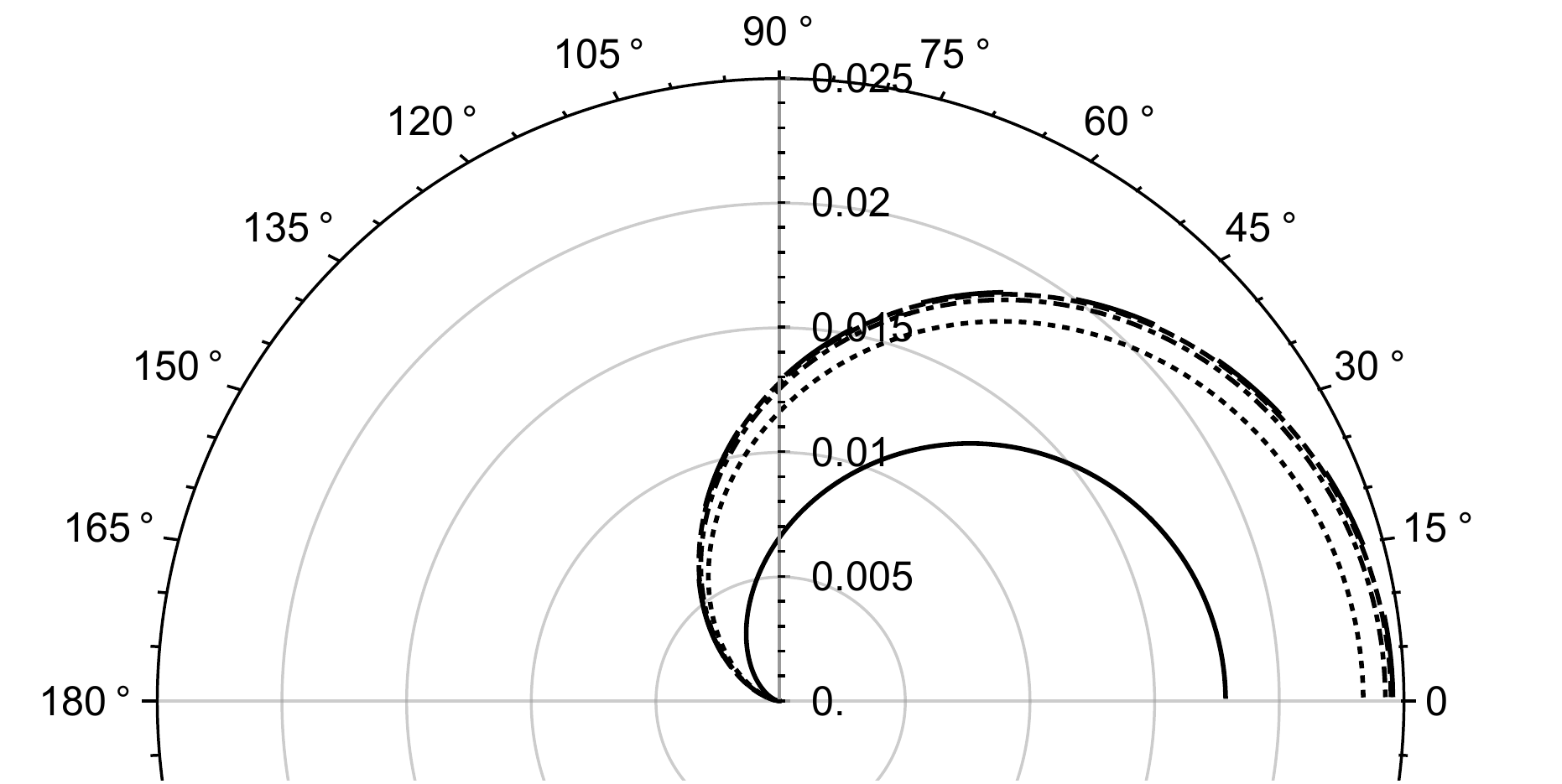}
\caption{$k_{1}=1$.}
\end{subfigure}
\hfill
\begin{subfigure}[b]{0.49\linewidth}
 \centering
 \includegraphics[width=1.1\textwidth]{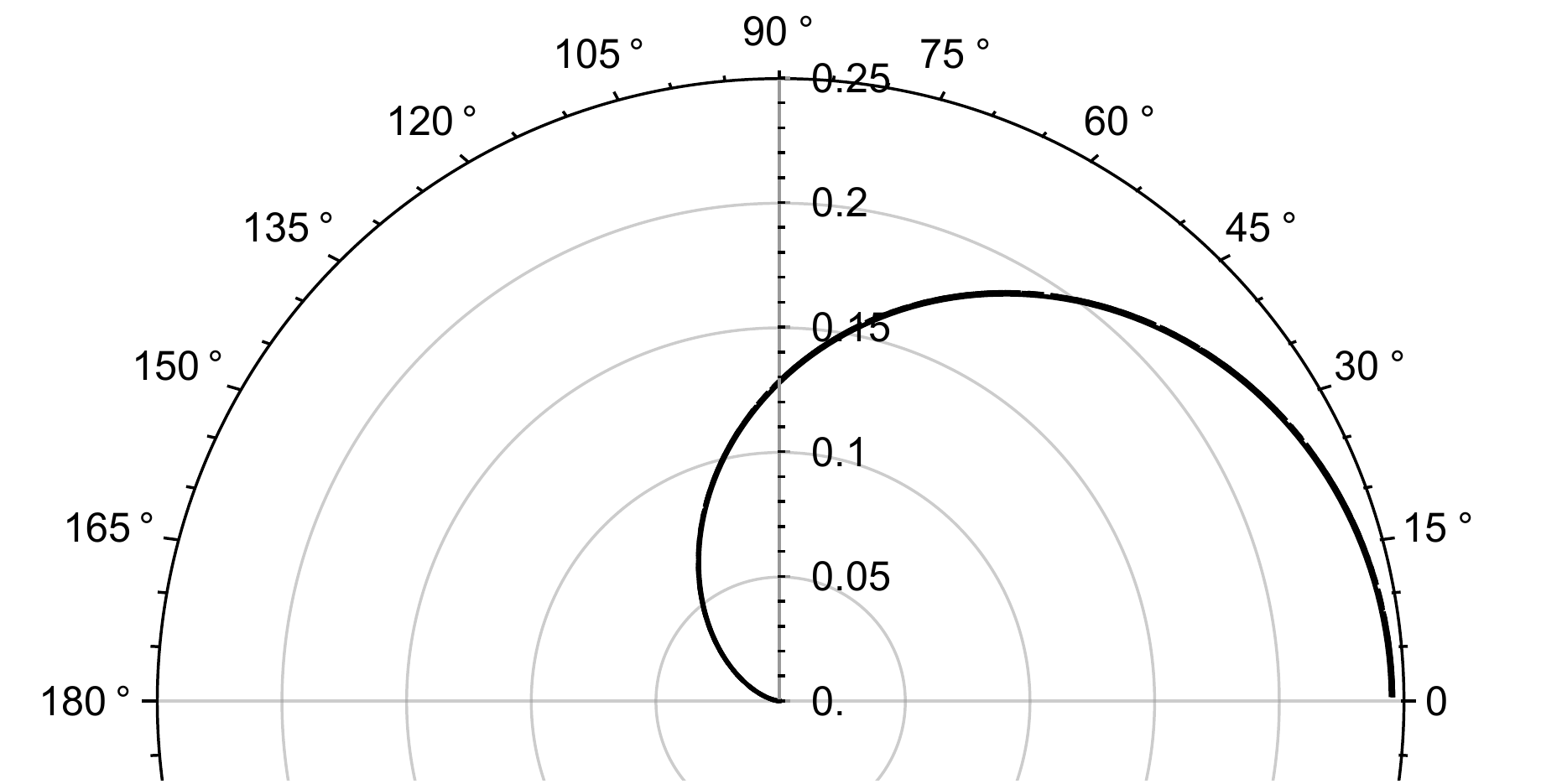}
\caption{$k_{1}=0.1$.}
\end{subfigure}
\caption{Polar plot of the spanwise average directivity, $D_{a}(r,\theta)$ as given by \eqref{eq:avedirect}, for case b) (periodic channel) with $r=10$, $M=0.3$, $k_{3}=0$. Large dashed $c=0$; dashed $c=0.5$; dot-dashed $c=1$; dotted $c=2$, solid $c=5$.}
\label{fig:periodic}
\end{figure}

We now consider the spanwise average directivity, $D_{a}(r,\theta)$, for the periodic case b). In Figure \ref{fig:periodic} we plot $D_{a}(r,\theta)$ over the same range of frequency and serration height as Figure \ref{fig:sweepcompare} when $k_{3}=0$. We see great similarity between the two figures since for low and mid-range frequencies only the $0$th order mode propagates to the far field, and these contributions are the same when $k_{3}=0$ (they yield the same $E_{0}$ coefficients). For higher frequencies (illustrated by $k_{1}=10$), or when $k_{3}\neq0$ we expect to see different behaviour between case a) and b) as the $E_{n}$ coefficients contributing to the far-field acoustics are no longer identical.

\begin{figure}
 \centering
\begin{subfigure}[b]{0.49\linewidth}
 \centering
 \includegraphics[width=1.1\textwidth]{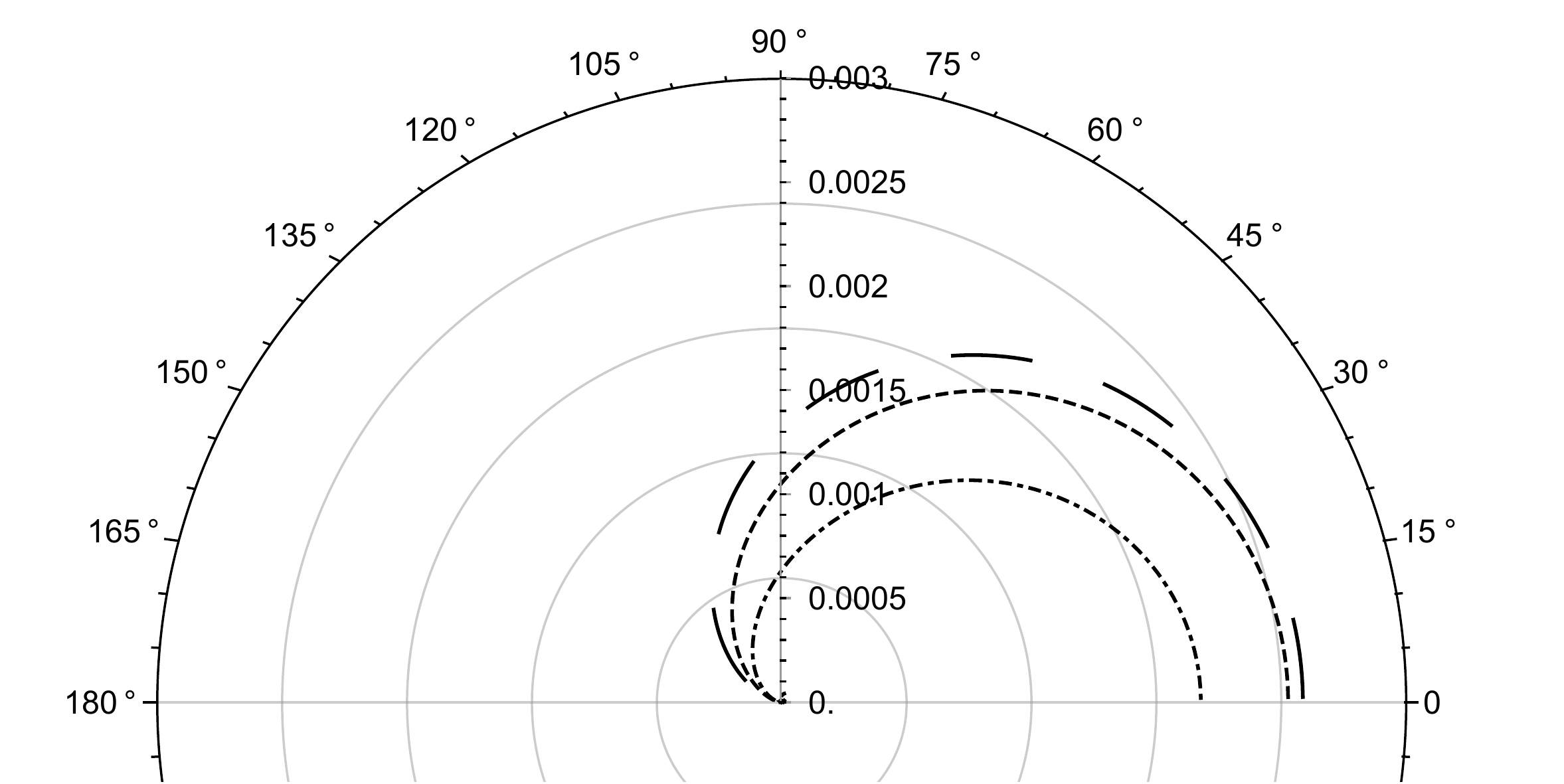}
\caption{$k_{1}=10$, $k_{3}=1$, Case a).}\label{fig:k3n0a}
\end{subfigure}
\hfill
\begin{subfigure}[b]{0.49\linewidth}
 \centering
 \includegraphics[width=1.1\textwidth]{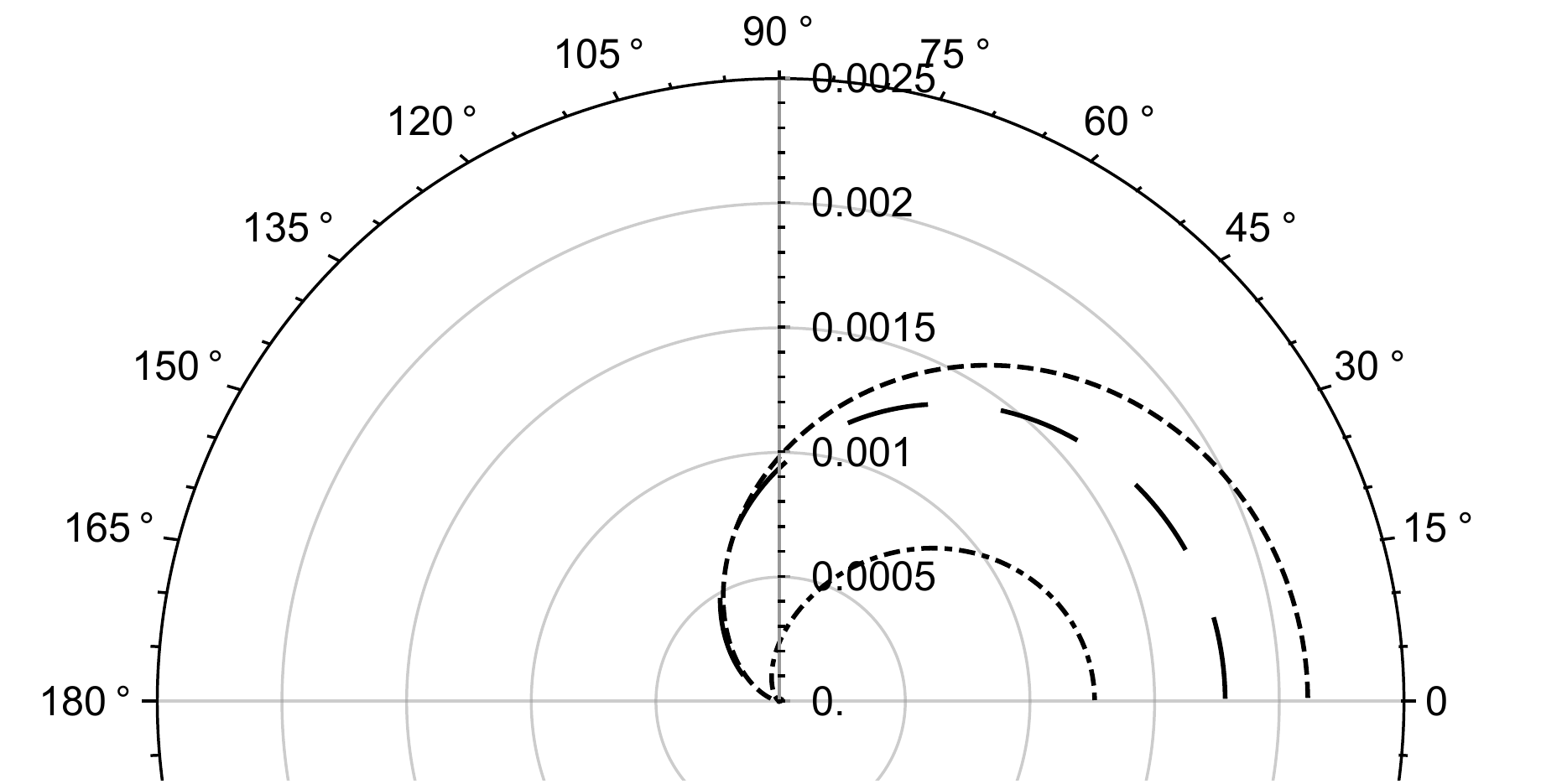}
\caption{$k_{1}=10$, $k_{3}=1$, Case b).}\label{fig:k3n0b}
\end{subfigure}
\\
\vspace{0.05\linewidth}
\hspace{-20pt}
\begin{subfigure}[b]{0.49\linewidth}
\centering
\includegraphics[width=1.3\textwidth]{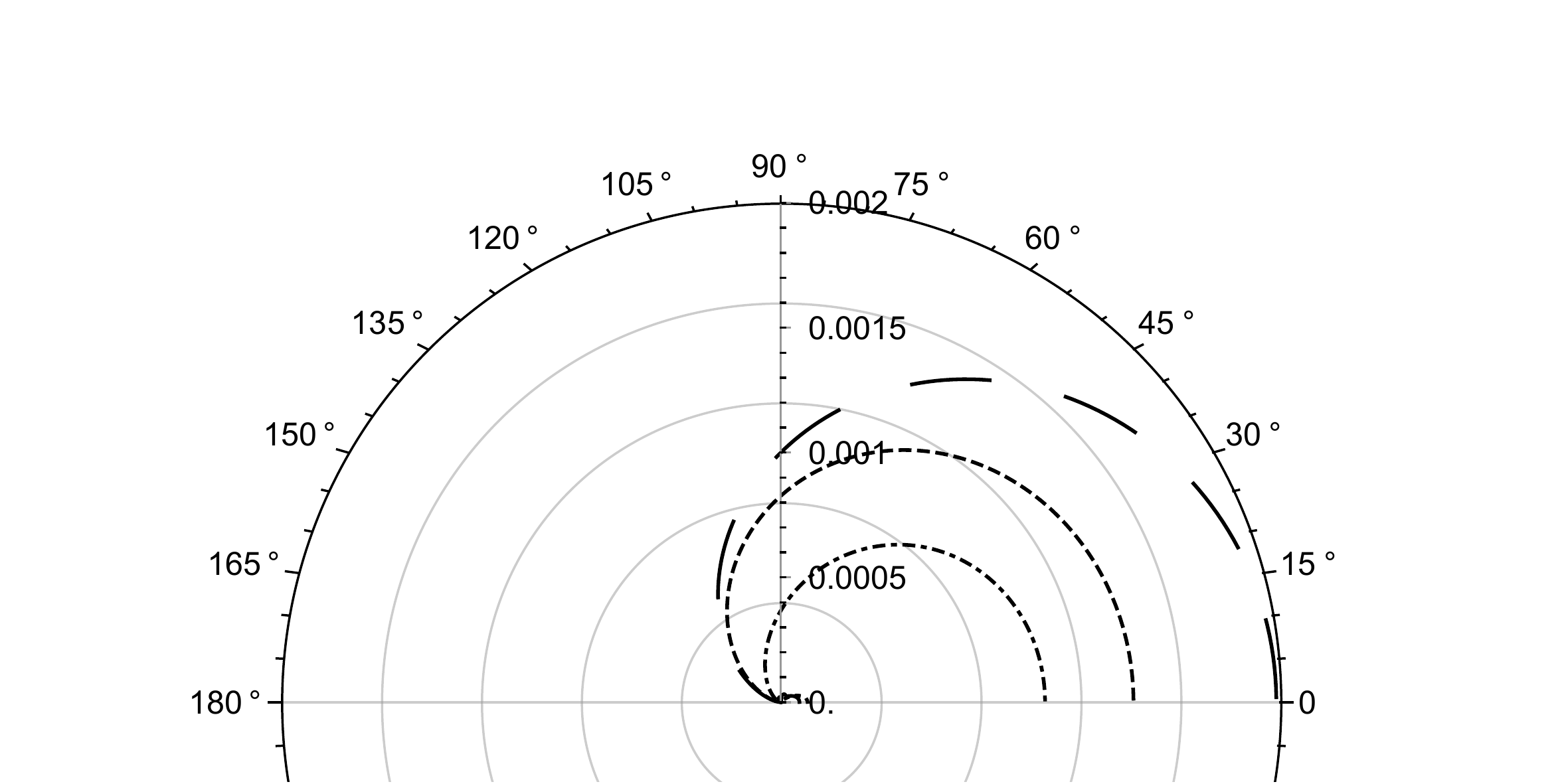}
\caption{$k_{1}=10$, $k_{3}=5$, Case a).}\label{fig:k3n0c}
\end{subfigure}
\hfill
\begin{subfigure}[b]{0.49\linewidth}
 \centering
 \includegraphics[width=1.1\textwidth]{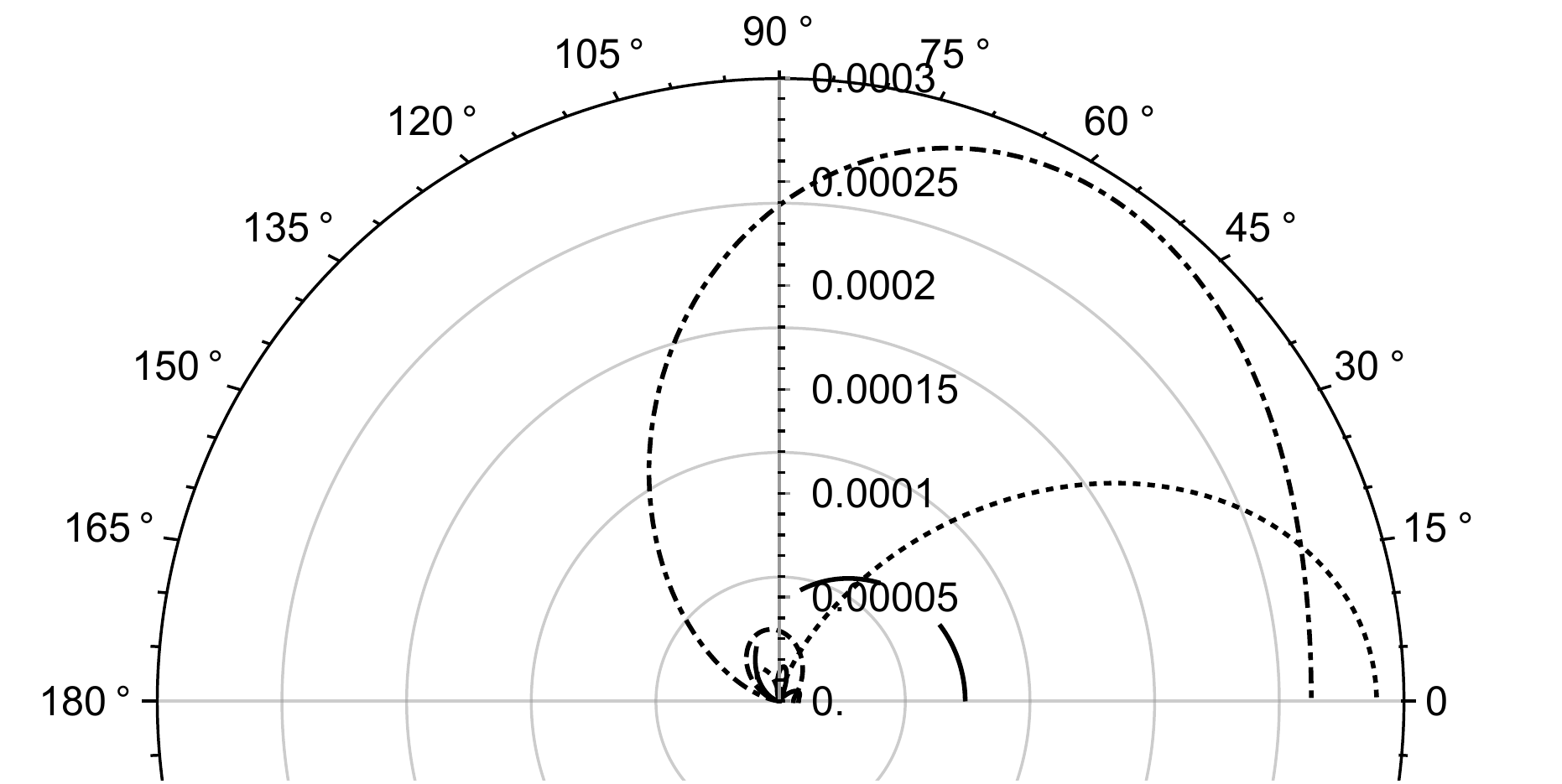}
\caption{$k_{1}=10$, $k_{3}=5$, Case b).}\label{fig:k3n0d}
\end{subfigure}
\\
\vspace{0.05\linewidth}
\begin{subfigure}[b]{0.49\linewidth}
 \centering
 \includegraphics[width=1.1\textwidth]{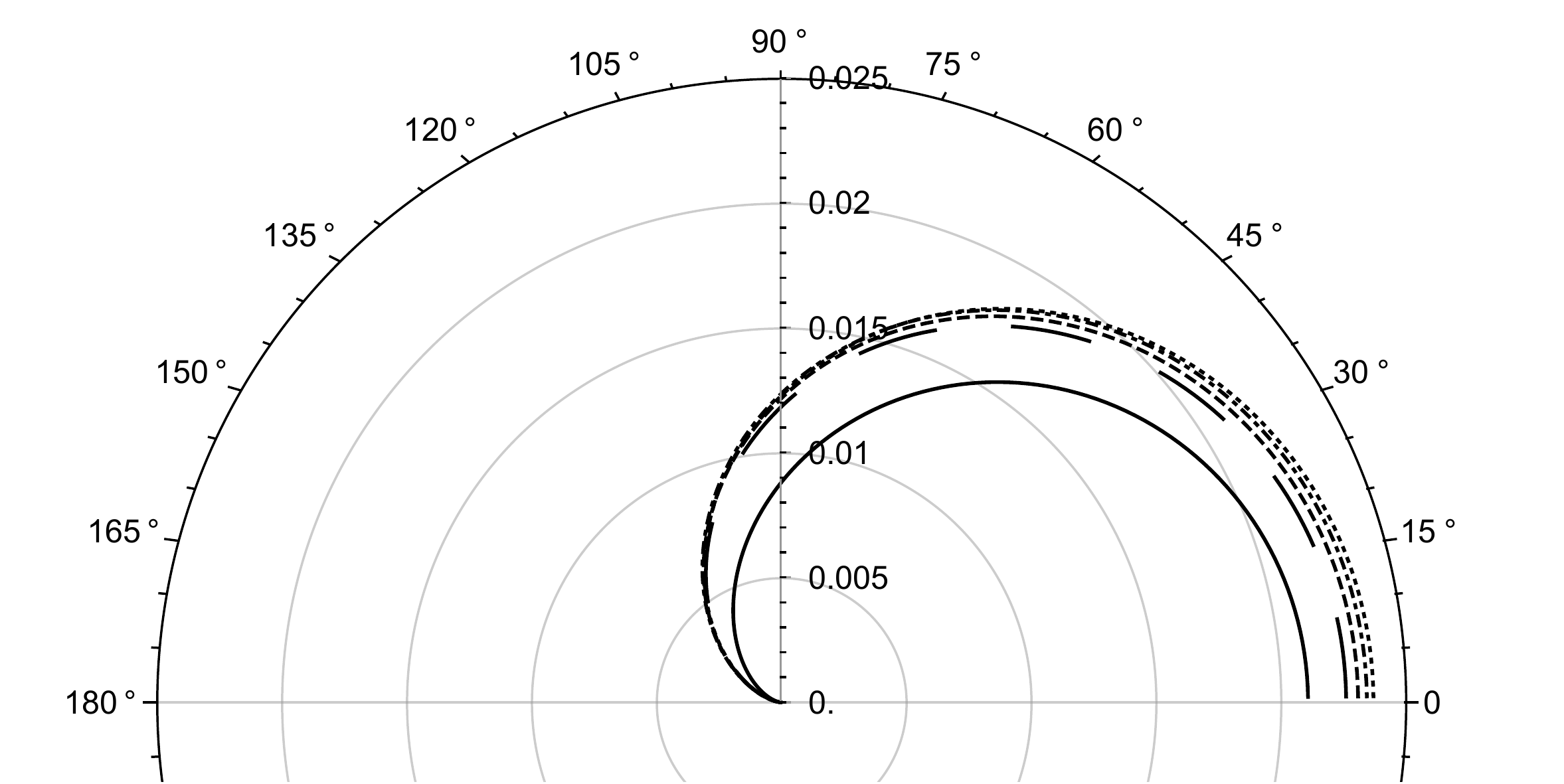}
\caption{$k_{1}=1$, $k_{3}=1$, Case a).}\label{fig:k3n0e}
\end{subfigure}
\hfill
\begin{subfigure}[b]{0.49\linewidth}
 \centering
  \includegraphics[width=1.1\textwidth]{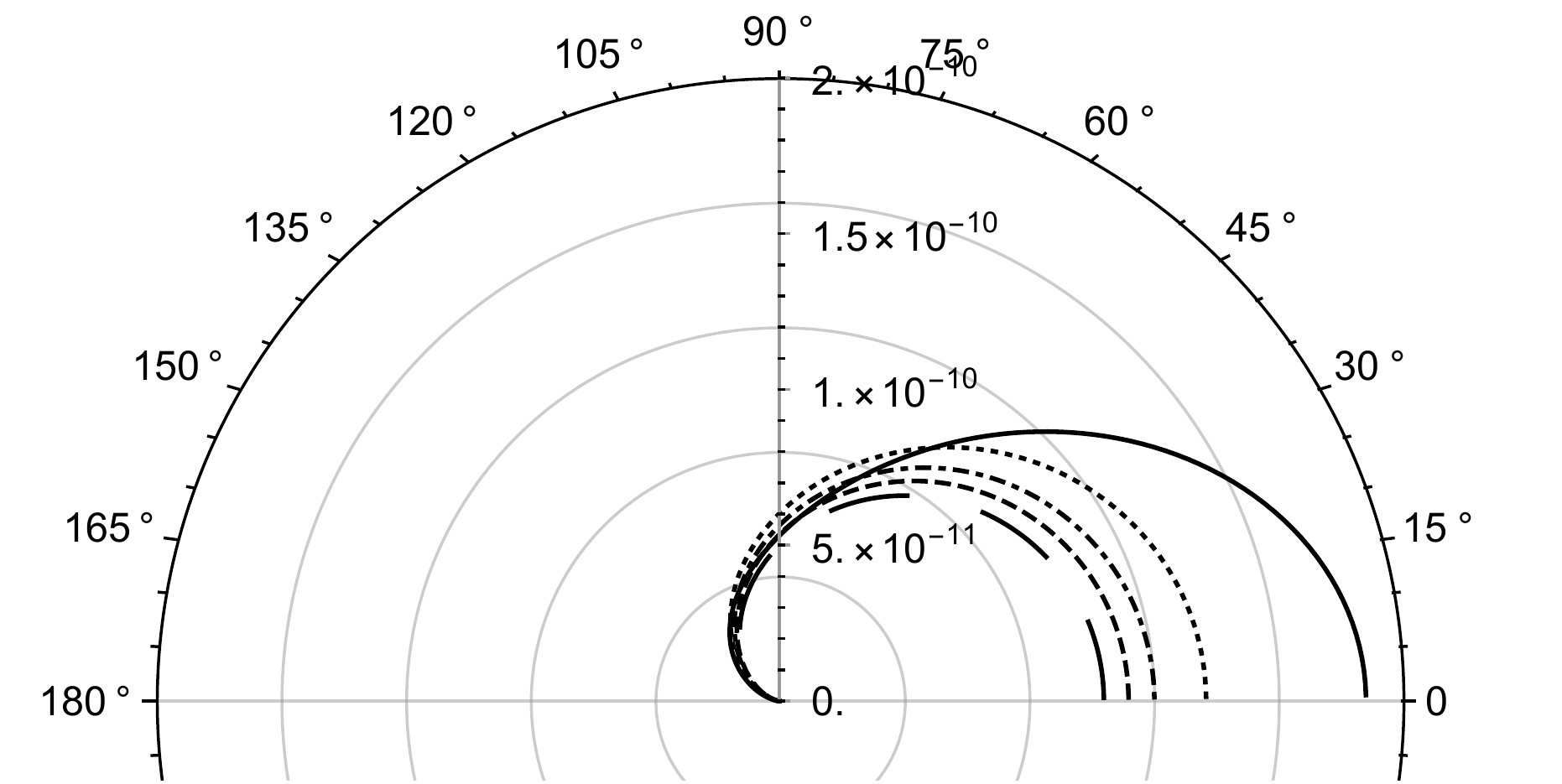}
\caption{$k_{1}=1$, $k_{3}=1$, Case b).}\label{fig:k3n0f}
\end{subfigure}
\\
\vspace{0.05\linewidth}
\begin{subfigure}[b]{0.49\linewidth}
 \centering
 \includegraphics[width=1.1\textwidth]{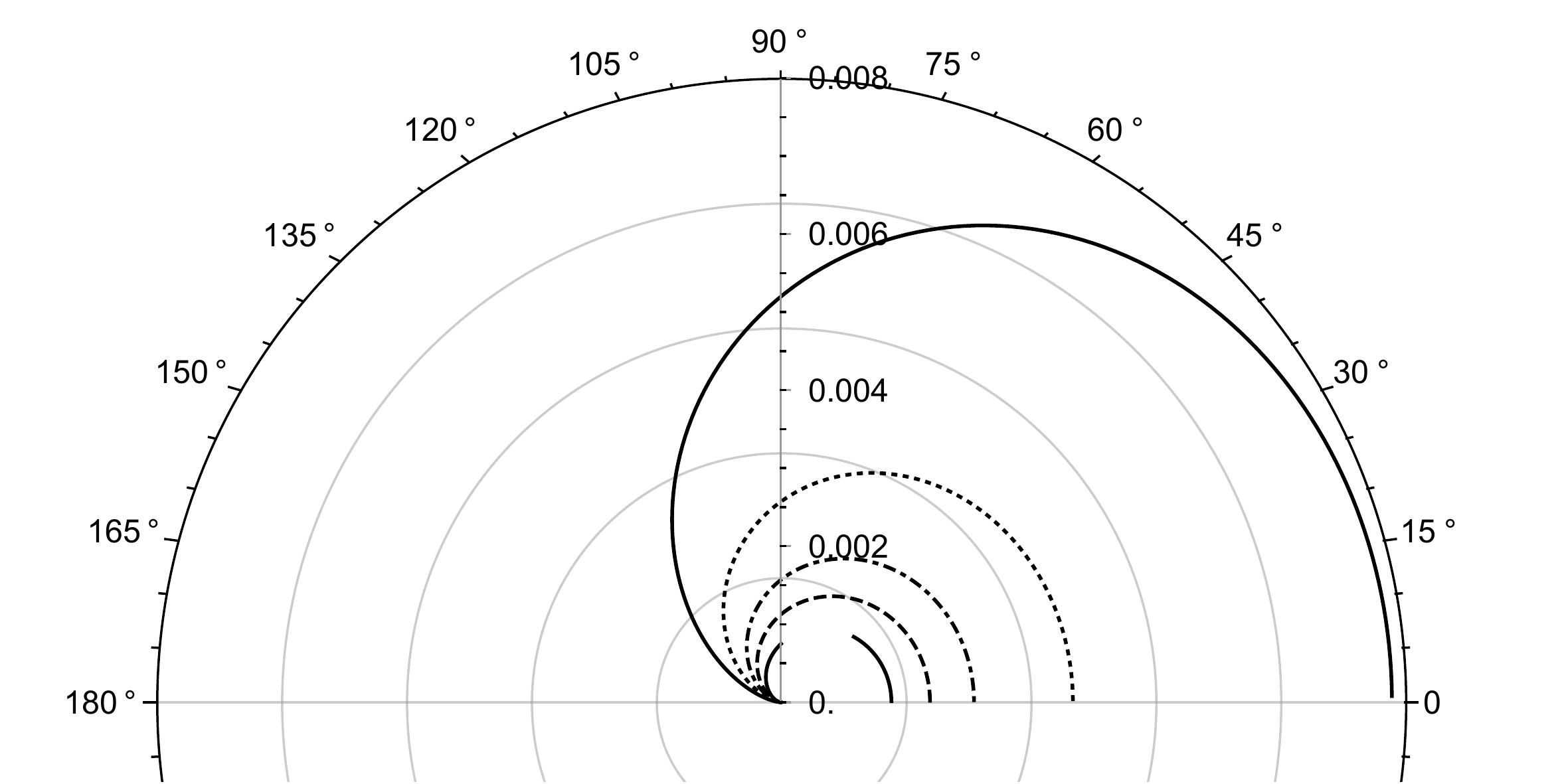}
\caption{$k_{1}=1$, $k_{3}=5$, Case a).}\label{fig:k3n0g}
\end{subfigure}
\hfill
\begin{subfigure}[b]{0.49\linewidth}
 \centering
\includegraphics[width=1.1\textwidth]{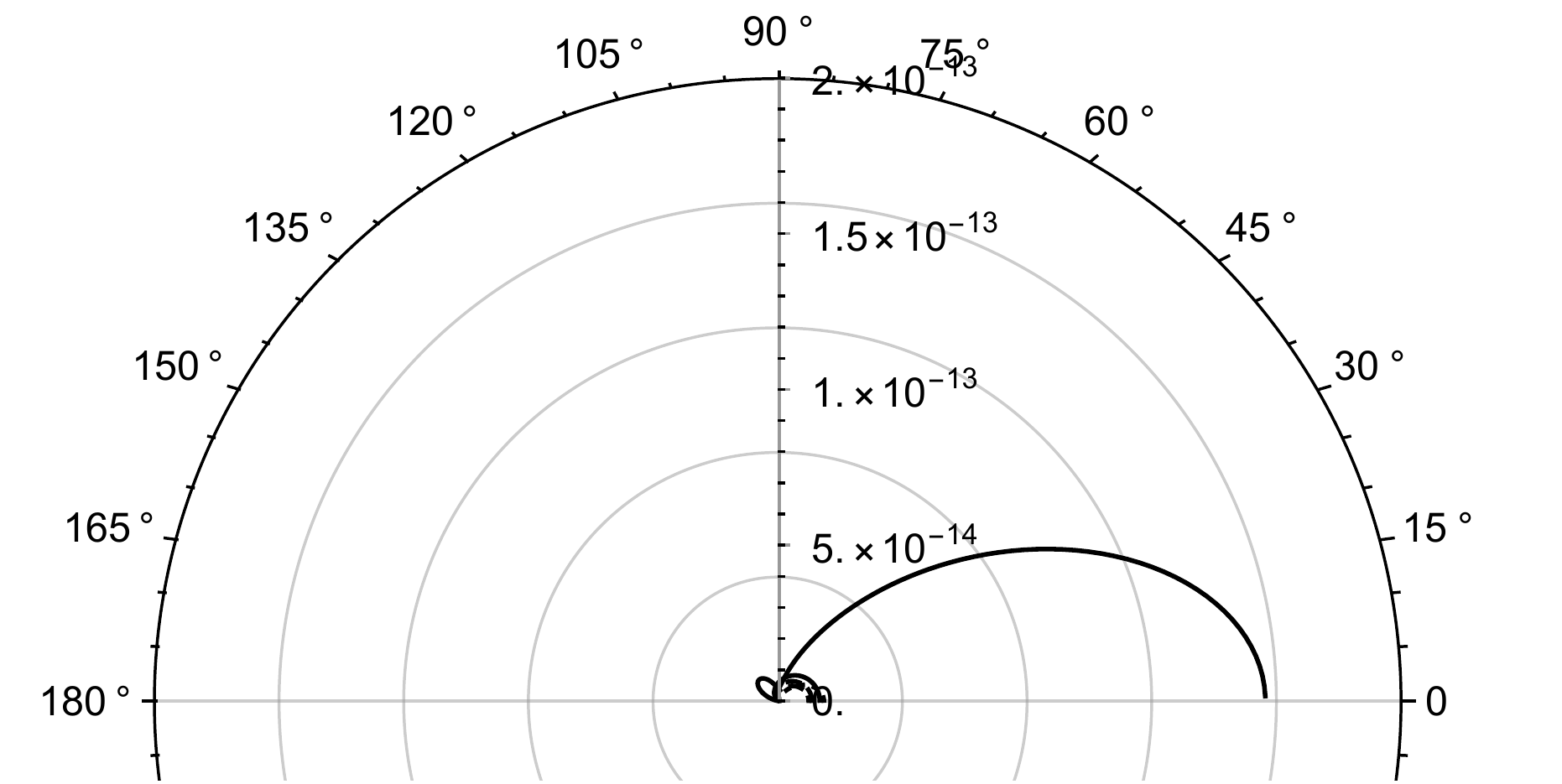}
\caption{$k_{1}=1$, $k_{3}=5$, Case b).}\label{fig:k3n0h}
\end{subfigure}
\caption{Polar plot of the spanwise average directivity, $D_{a}(r,\theta)$ as given by \eqref{eq:avedirect}, for $r=10$, $M=0.3$. Large dashed $c=0$; dashed $c=0.5$; dot-dashed $c=1$; dotted $c=2$, solid $c=5$.}
\label{fig:k3n0}
\end{figure}

In Figure \ref{fig:k3n0} we compare cases a) and b) when $k_{3}\neq0$. For the periodic boundary conditions of case b) a non-zero $k_{3}$ can cutoff all scattered frequencies $w_{n}$, but this is not true in the rigid walled case a). This cutoff is similar to a spanwise-infinite straight edge and is therefore not specifically a feature of the serration. Note when cutoff the results only include the first $6$ modes.
For the cases that are not cutoff, the periodic serrated edge (case b), shows the ability to increase noise versus a straight edge ($c=0$) for large $k_{1}$ and similar sized $k_{3}$ (Figure \ref{fig:k3n0d}) when $c\leq1$, however a reduction of noise occurs for $c\geq2$. The walled serration, case a), sees little noise reduction in similar circumstances (Figure \ref{fig:k3n0c}), but only exhibits a consistent noise increase with increasing $c$ when $k_{3}$ is significantly larger than $k_{1}$. We note for high $k_{1}$ and (relative) small $k_{3}$, such as Figure \ref{fig:k3n0a},\ref{fig:k3n0b}, the periodic serration results in a slightly greater reduction of far-field noise than the walled serration.

Overall we see serrations either in a rigid walled channel or periodic channel reduce gust interaction noise with increasing serration heights. Some increases of noise can also occur for low tip-to-root heights.

\subsection{Far-field PSD for periodic channel}\label{sec:PSD}

In the previous sections we have considered the far-field noise at fixed $k_{3}$ values. In this section we integrate over a spectrum of $k_{3}$ values to calculate the far-field power spectral density (PSD) and compare to experimental measurements from \citet{sn15}. We use an upstream spectrum defined by
\begin{equation}
 \Phi^{(\infty)}(k_1,k_3)=\frac{k_1^2/k_e^2+k_3^2/k_e^2}{\left(1+k_1^2/k_e^2+k_3^2/k_e^2\right)^{7/3}},
\end{equation}
where 
\begin{equation*}
 k_{e}=\frac{\sqrt{\upi}\,\Gamma(5/6)}{\Gamma(1/3)L_{t}},
\end{equation*}
 and $L_{t}=0.6$ is the non-dimensionalised lengthscale of turbulence.
The far-field PSD is thus defined as
\begin{equation}
 \text{PSD}=\int_{-\infty}^{\infty}|p(r,\theta,z)|^{2}\Phi^{(\infty)}(k_{1},k_{3})dk_{3}.
\end{equation}
Large $k_{3}$ values have been seen to cut off the scattered field, thus in practise when numerically evaluating this quantity from the analytic expression for $p$, we integrate only over a finite range of $k_{3}$ corresponding to cut on modes.

\begin{figure}
\centering
\begin{subfigure}[b]{0.49\linewidth}
 \centering
 \includegraphics[width=1\textwidth]{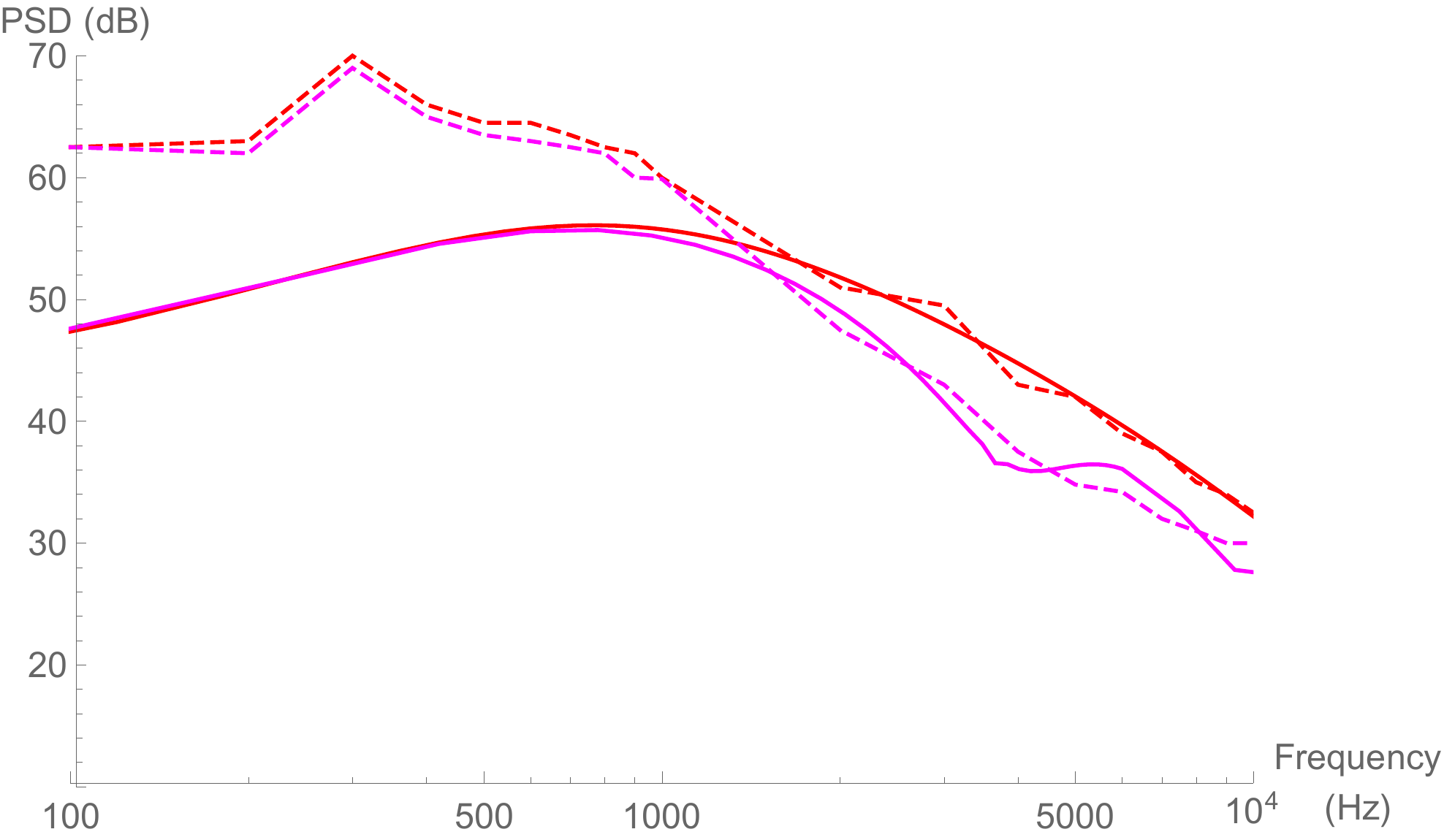}
\caption{Straight edge $c=0$ red, and serrated edge $c=2$ pink.}
\end{subfigure}
\hfill
\begin{subfigure}[b]{0.49\linewidth}
 \centering
 \includegraphics[width=1\textwidth]{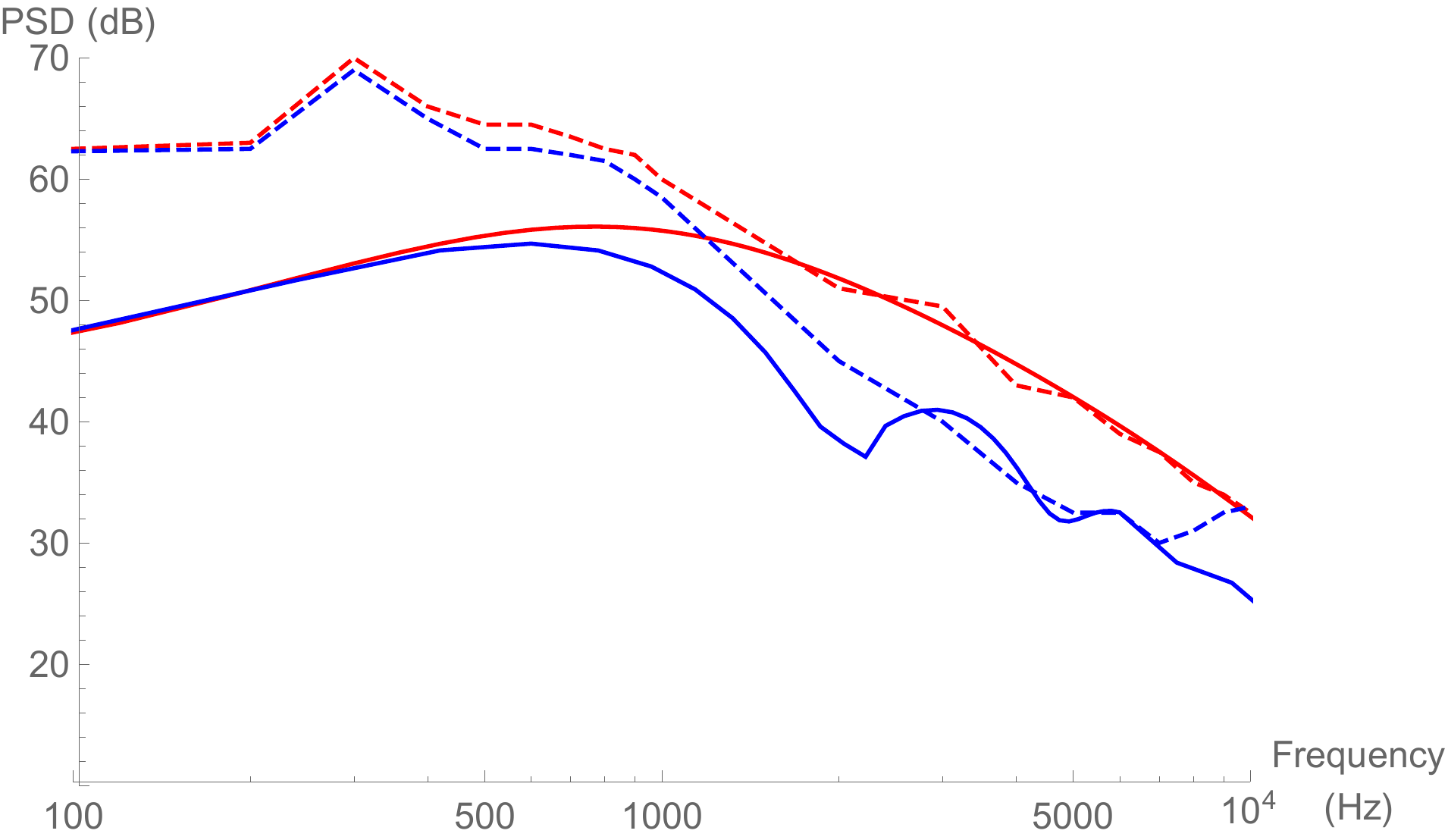}
\caption{Straight edge $c=0$ red, and serrated edge $c=4$ blue.}
\end{subfigure}
\caption{Comparison of far-field PSD for two serrated leading edges. Solid lines give the analytic results, dashed lines give the experimental measurements.}
\label{fig:PSD}
\end{figure}

In Figure \ref{fig:PSD} we compare the analytic predictions for the PSD to experimental measurements from \citet{sn15}, taken at $\theta=\upi/2$, and mid-span $z=0.5$. The serrations correspond to $c=2,4$. We see very good agreement between the analytic and experimental results at mid-range frequencies, indicating that the simple analytic model is capturing all of the key physics behind the noise reductions. We do not expect good agreement at low frequencies due to the dominance of jet noise in the experimental measurements. Similarly at very high frequencies trailing-edge noise dominates the experimental measurements thus we do not expect agreement for frequencies beyond around $10^4$ Hz.

We note that whilst the straight-edge analytic PSD is non-oscillatory (as expected for a single scattering location, the leading edge), the serrated PSD results oscillate with increasing frequency. This indicates the destructive interference effect discussed at fixed $k_{3}$ values for the individual modes, $B_{n}$, remains true for a spectrum of $k_{3}$ values.

\subsection{Noise reduction mechanisms}\label{sec:reds}

The mathematical solution allows us to see that the reduction occurs due to two key properties: first the interference between terms in the modal expansion coefficients, $E_{n}(-w_{n}\cos\theta)$, which depend on both wavenumber components $k_{1}$ and $k_{3}$, and the tip-to-root ratio of the serration. Generally destructive interference occur, however constructive interference can also occur at fixed $k_{3}$ values and in these cases we see the increase of overall far-field noise. Second; a redistribution energy towards higher modes for large serration heights. 

We illustrate these two properties for periodic case b) by considering the modal expansion terms (for simplicity when $k_{3}=0$) given by
\begin{equation}
 E_{n}=\frac{4 (-1)^{n}s\,\sin\left(\frac{s+2n\upi}{4}\right)}{(s+2n\upi)(s-2n\upi)},
\label{eq:Enk30}
\end{equation}
where $s=\gamma(\kappa+\lambda)$. When evaluating in the far field, we set $\lambda=-w_{n}\cos\theta$, hence $s=k_{1}c\beta^{-2}(1-M \cos\theta)$. The interference is clearly represented by the oscillatory sine term, which oscillates with varying (reduced) frequency, $k_{1}$, and with varying tip-to-root ratio, $c$.

For small serration heights, $s\to0$, $E_{0}\to1$ and $E_{n\neq0}\to0$ (which is what would be obtained if we did a modal expansion when $F(z)=0$ for a straight edge). For large $s$, when $s\gg m$, $E_{m}\to0$, but for $s=O(m)$, $E_{m}\neq0$, thus the low $E_{n}$ coefficients tend to zero whilst the higher modes \emph{which do not propagate to the far field} do not tend to zero. Thus there is a redistribution of acoustic energy towards cutoff modes hence the far-field noise decreases.

This simple formula for the modal terms, \eqref{eq:Enk30}, also allows us to predict the rate of noise reduction for increasing tip-to-root ratio. In particular, for large serration heights, $c\gg 1$, $E_{n}\sim c^{-1}$, therefore the far-field noise reduction in decibels is logarithmic, $\sim\log_{10}(c)$, which is as predicted via the numerical and experimental investigations of \citet{sn15}.
This additional noise-reduction mechanism (the redistribution of energy between modes) ensures that even if a constructive interference is occurring in the modal expansion terms, the overall combination of the two mechanisms leads to one observing an overall decrease of noise in the far field (except at low values of $c$ for which the redistribution is less effective).

The modal coefficients arise due to the expansion of the normal velocity on the leading edge. If the leading edge were straight (spanwise infinite), the expansion of a single frequency normal velocity would be into a Fourier series with a single viable mode, $\e^{i k_{3}\zeta}$, as if we just factored the $\zeta$ dependence from the problem. For a serrated edge, the term $\e^{i\kappa\gamma F(\zeta)}$ must be decomposed into the modal basis which is now not a simple Fourier series, but also dependent on $\gamma F(\zeta)$, and requires multiple modes each of which have coefficients dependent on $\gamma$ and $k_{3}$ allowing for individual modes to have a destructive interference and a redistribution of energy among different modes. 

\subsection{Numerical results}\label{sec:numerics}
In addition to the comparison of the analytic predictions to experimental results (figure \ref{fig:PSD}), here we compare the analytic results, for convective gust plate interaction, to numerical results for line vortex plate interaction. Whilst we anticipate the different incident fields to produce subtly different final results, it is expected that given both incident fields represent upstream turbulence, they will yield qualitatively similar far-field predictions. This will show a range of different models of turbulence can be used to qualitatively predict the same effects. Different models of incoming turbulence are beneficial as different approaches (e.g. analytical or numerical) are best suited to different incident fields.

%%%%%%%%%%%%%%%%%%%%

\subsubsection{Description of the numerical solution approach}
The current numerical solutions are achieved by using the same approach published recently by \cite{Turner2017}. The only adjustments made for this particular work are to change the leading-edge serration geometry from a sinusoidal to a sawtooth; and, to implement the periodic condition on the spanwise boundaries. The current computation employs full three-dimensional compressible Euler equations in a conservative form transformed onto a generalised coordinate system:
\begin{equation}\label{eq:Euler}
\frac{\partial}{\partial t}\left(\frac{\boldsymbol{Q}}{J}\right)+\frac{\partial}{\partial\xi_i}\left(\frac{\boldsymbol{F}_j}{J}\frac{\partial\xi_i}{\partial x_j}\right)=-\frac{a_\infty}{L_c}\frac{\boldsymbol{S}}{J},
\end{equation}
where $a_\infty$ is the ambient speed of sound; $L_c$ is a characteristic length scale ($L_c=10$, i.e. 10 times the serration wavelength in this paper); and, the indices $i=1,2,3$ and $j=1,2,3$ denote the three dimensions. In \eqref{eq:Euler}, the conservative variable and flux vectors are given by
\begin{equation}\label{eq:Flux}
\left.
\begin{gathered}
\boldsymbol{Q}=[\rho,\rho u,\rho v,\rho w,\rho e_\text{t}]^T,\\
\boldsymbol{F}_j=[\rho u_j,(\rho uu_j+\delta_{1j}p),(\rho vu_j+\delta_{2j}p),(\rho wu_j+\delta_{3j}p),(\rho e_\text{t}+p)u_j]^T,
\end{gathered}
\right\}
\end{equation}
where $\xi_i=\{\xi,\eta,\zeta\}$ are the generalised coordinates, $x_j=\{x,y,z\}$ are the Cartesian coordinates, $\delta_{ij}$ is the Kronecker delta, $u_j=\{u,v,w\}$, $e_\text{t}=p/[(\gamma-1)\rho]+u_ju_j/2$ and $\gamma=1.4$ for air. In the current setup, $\xi$, $\eta$ and $\zeta$ are body fitted coordinates along the grid lines in the streamwise, vertical and lateral directions, respectively. The Jacobian determinant of the coordinate transformation (from Cartesian to the body fitted) is given by $J^{-1}=|\partial(x,y,z)/\partial(\xi,\eta,\zeta)|$ \citep{Kim2002AIAA}. The extra source term $\boldsymbol{S}$ on the right-hand side of \eqref{eq:Euler} is arranged to implement a non-reflecting sponge condition which is detailed in \citet{Kim2010a,Kim2010b}.

The governing equations given above are solved by using high-order accurate numerical methods specifically developed for aeroacoustic simulations on structured grids as discussed in \citet{Turner2017}. 
% The flux derivatives in space are calculated based on fourth-order pentadiagonal compact finite difference schemes with seven-point stencils \citep{Kim2007}. Explicit time advancing of the numerical solution is carried out by using the classical fourth-order Runge-Kutta scheme with the CFL number of 0.95. Numerical stability is maintained by implementing sixth-order pentadiagonal compact filters for which the cutoff wavenumber (normalised by the grid spacing) is set to $0.85\pi$ \citep{Kim2010}. In addition to the sponge layers used, characteristics-based non-reflecting boundary conditions \citep{Kim2000} are applied at the far-boundaries in order to prevent any outgoing waves from returning to the computational domain. Periodic or slip-wall (no penetration) conditions are used for the spanwise boundary planes as indicated earlier. The slip-wall boundary conditions are also used on the aerofoil surface \citep{Kim2004}, which is extended downstream (all the way down to the exit boundary) to mimic the semi-infinite chord of the flat-plate aerofoil.
The computational domain is a cuboid that covers $4.5L_c$ (including the sponge layers) in the upstream, vertical and downstream directions from the mean leading edge position. The spanwise length is equal to the serration wavelength. 
The domain is filled with a structured grid that is uniform in the majority of the domain except in the sponge zone and the local area nearest to the leading edge. 
% The total grid cell count is $n_\xi \times n_\eta \times n_\zeta=600 \times 600 \times 64=23,040,000$ where $n_\xi$, $n_\eta$ and $n_\zeta$ are the number of cells in the streamwise, vertical and lateral/spanwise directions, respectively. The smallest cells are positioned around the leading edge where 
% $\rmDelta x_{\min}=\rmDelta y_{\min}=L_c/200$ and $\rmDelta z_{\min}=L_c/640$. 
A high grid density is maintained in the acoustic field in order to accurately capture the high-frequency components (typically 10 cells per acoustic wavelength in the upstream direction at the frequency of $k_1=10$ in this paper). In total, 12.8 million grid cells (448$\times$448$\times$64) are used in the current numerical simulations.

The computation is parallelised via domain decomposition and message passing interface (MPI) approaches. The compact finite difference schemes and filters used are implicit in space due to the inversion of pentadiagonal matrices involved, which requires a precise and efficient technique for the parallelisation in order to avoid numerical artifacts that may appear at the subdomain boundaries. A recent parallelisation approach based on quasi-disjoint matrix systems \citep{Kim2013} offering super-linear scalability is used in the present paper. The entire domain is decomposed and distributed onto 392 separate processor cores ($14 \times 14 \times 2$ in the streamwise, vertical and spanwise directions, respectively). The parallel computation has successfully been carried out in the IRIDIS-4 computer cluster at the University of Southampton.
\subsubsection{Prescribed spanwise vortex model}
The current numerical simulation employs a spanwise vortex model prescribed as an initial condition, instead of a harmonic vortical gust used for the analytical solution. This approach has an advantage of having a wide range of frequency responses in a single simulation as opposed to running multiple separate simulations each for a single-frequency gust. The vortex model is based on a vector potential function:
\begin{equation}\label{eq:vectorpotential}
\vec{\psi}(\xx)=\frac{\epsilon}{2\pi}a_\infty R \left(\frac{r}{R}\right)^\frac{3}{2}\exp\left[1-\frac{1}{2}\left(\frac{r}{R}\right)^2\right]\vec{e}_z \quad\mathrm{with}\quad r=\sqrt{(x-x_0)^2+y^2},
\end{equation}
where $R$ is a representative length scale of the vortex and $\vec{e}_z$ is a unit vector in the spanwise direction. The velocity field is created by taking the curl of the vector potential, which provides a divergence-free initial condition:
\begin{align}\label{eq:vortexmodel}
\vec{u}(\xx)&=\nabla\times \psi(\xx)=\psi(\xx)\left\{M_\infty+\frac{y}{R}\sigma(\xx)\;,\;-\frac{x-x_0}{R}\sigma(\xx)\,,0\right\},\\
\sigma(\xx)&=1-\frac{3R^2}{(x-x_0)^2+y^2},
\end{align}
where $x_0$ is the initial streamwise position of the vortex. The pressure and density are determined by assuming an isentropic flow with its total enthalpy conserved:
\begin{equation}\label{eq:vortex-pressure}
\rho(\xx)=\rho_\infty\left[1-\frac{\gamma-1}{2}\left(\frac{\psi(\xx)}{a_\infty R}\right)^2\right], \quad p(\xx)=p_\infty\left[\frac{\rho(\xx)}{\rho_\infty}\right]^\gamma.
\end{equation}
The subscript `$\infty$' denotes the free-stream condition. The free-stream Mach number is set to $M_\infty=0.3$ for the current simulations. The free parameters $R$ and $\epsilon$ in \eqref{eq:vectorpotential} are set to $R=0.05L_c$ and $\epsilon=0.08$, which results in the largest vertical velocity perturbation to reach 5\% of the free-stream velocity.

%%%%%%%%%%%%%%%%%%%%%%%%%%%
\subsubsection{Comparison of numerical and analytical results}

A key difference between the analytic and numeric approaches is in the modelling of a simple incident turbulence component. Gusts are ideal for the analytical model since the normal velocity can be Fourier transformed analytically and its result is very amenable to the Wiener-Hopf factorisation process. A line vortex does not permit such analytic progress, however is simple to numerically implement. We therefore expect to see differences between the numerical and analytical results for fixed frequencies and serration heights, however as both are good models of turbulence interaction, we expect the key features to be the same as we alter either frequency or serration height, hence the same key mechanisms of noise reduction to be in action.

\begin{figure}
\centering
\begin{subfigure}[b]{0.49\linewidth}
 \centering
 \includegraphics[width=\textwidth]{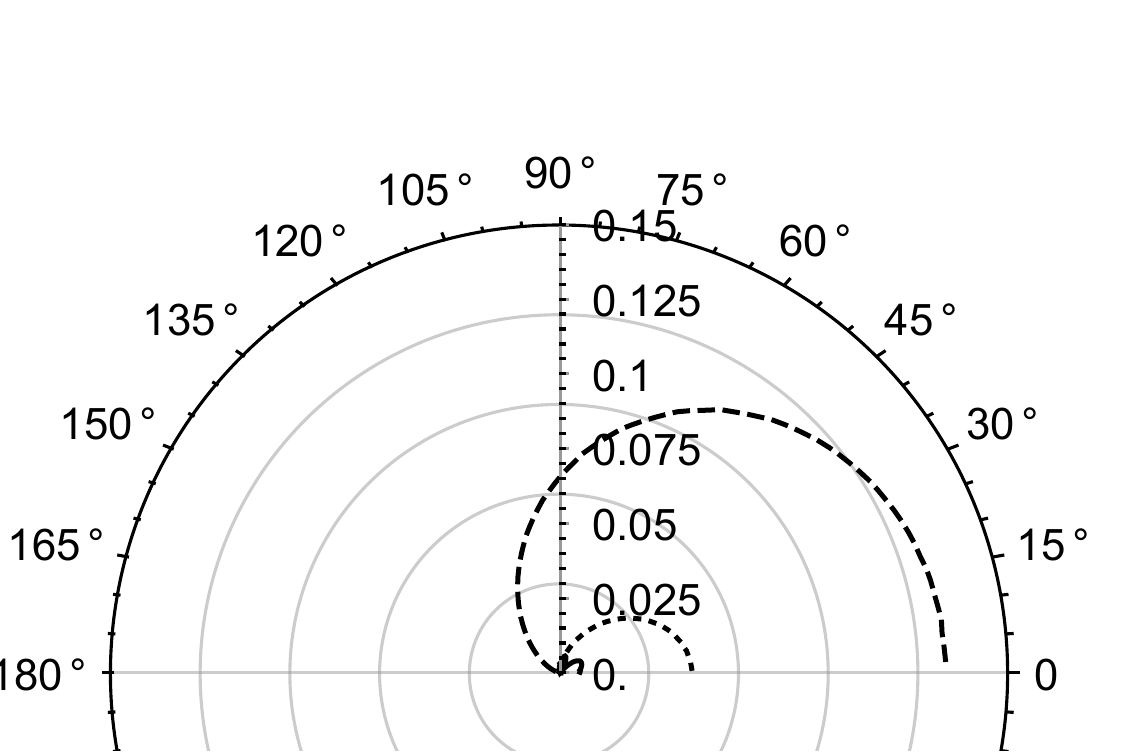}
\caption{Case a)}
\end{subfigure}
\hfill
\begin{subfigure}[b]{0.49\linewidth}
\hspace{-20pt}
\includegraphics[width=1.3\textwidth]{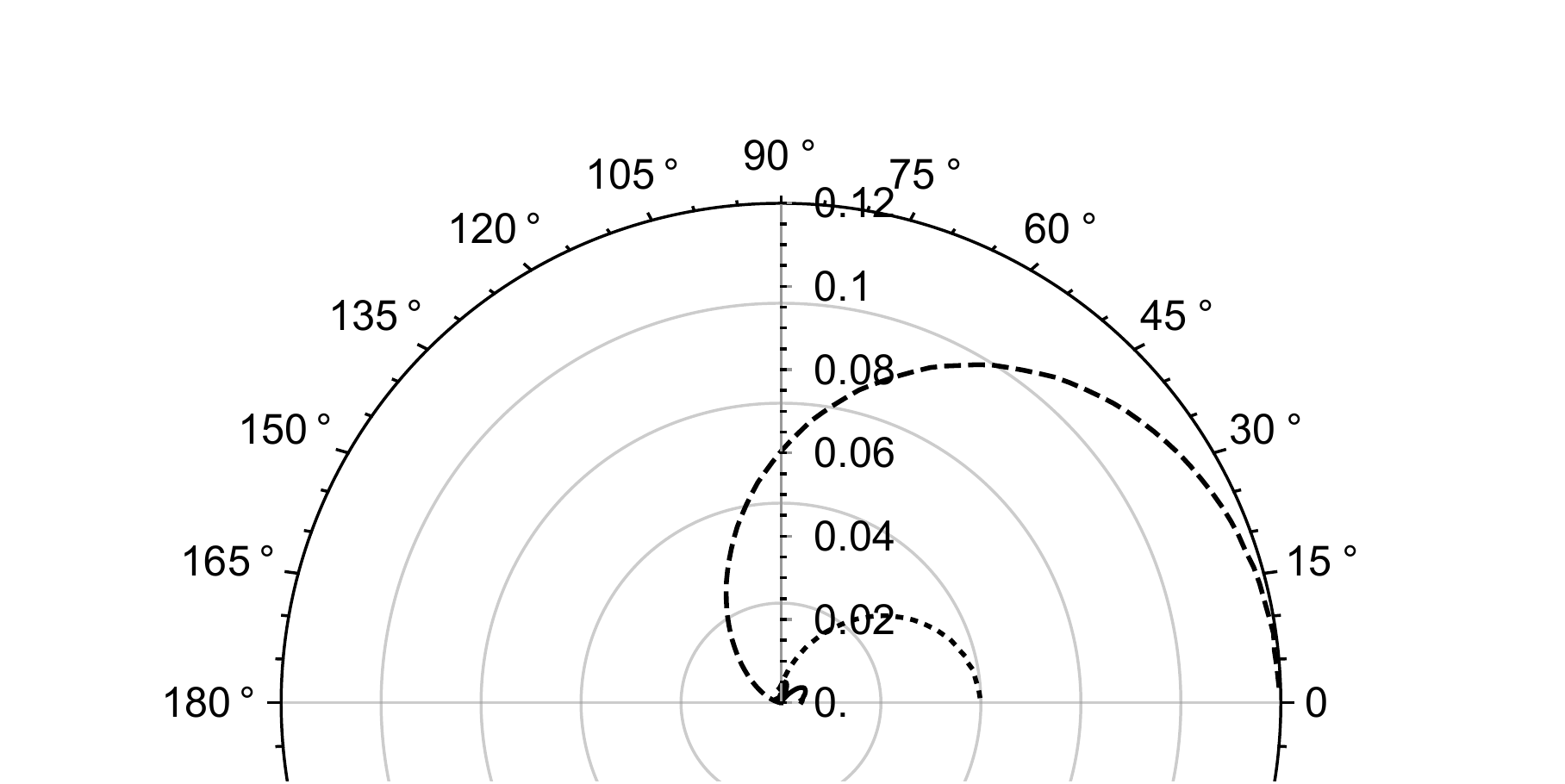}
\caption{Case b)}
\end{subfigure}
\caption{Polar plot of the spanwise average directivity multiplied by frequency, $k_{1} D_{n}(r,\theta)$, found numerically with incident vortex and $r=10$, $M=0.3$, $k_{3}=0$, $c=4$. Dashed $k_1=1$; dotted $k_1=5$; solid $k_1=10$.}
\label{fig:numericalAB}
\end{figure}

We first compare the numerical results from case a) to those from case b) in Figure \ref{fig:numericalAB}, for fixed serration height with $c=4$, for varying frequencies, $k_{1}=1,5,10$. We define $D_{n}(r,\theta)$ as the equivalent span-averaged directivity to the analytically defined $D_{a}(r,\theta)$, and multiply by the frequency to ensure details at $k_{1}=10$ can be viewed.
It is clear both case a) and case b) show very similar acoustic directivities in Figure \ref{fig:numericalAB} for $k_{1}=1,5$, which is as predicted analytically by comparing the results in Figures \ref{fig:sweepcompare} and \ref{fig:periodic}. Through the analytic solution we are able to attribute this to the fact that at these frequencies there is only one scattered mode, and for zero spanwise wavenumber these modes are identical for cases a) and b).
For $k_{1}=10$ the directivities differ, also as predicted by the analytic results. Evident too in Figure \ref{fig:numericalAB} is the trend of increasing $k_{1}$ increasing the modulation of the far-field, which is found analytically.

\begin{figure}
\centering
\begin{subfigure}[b]{0.49\linewidth}
 \centering
 \includegraphics[width=\textwidth]{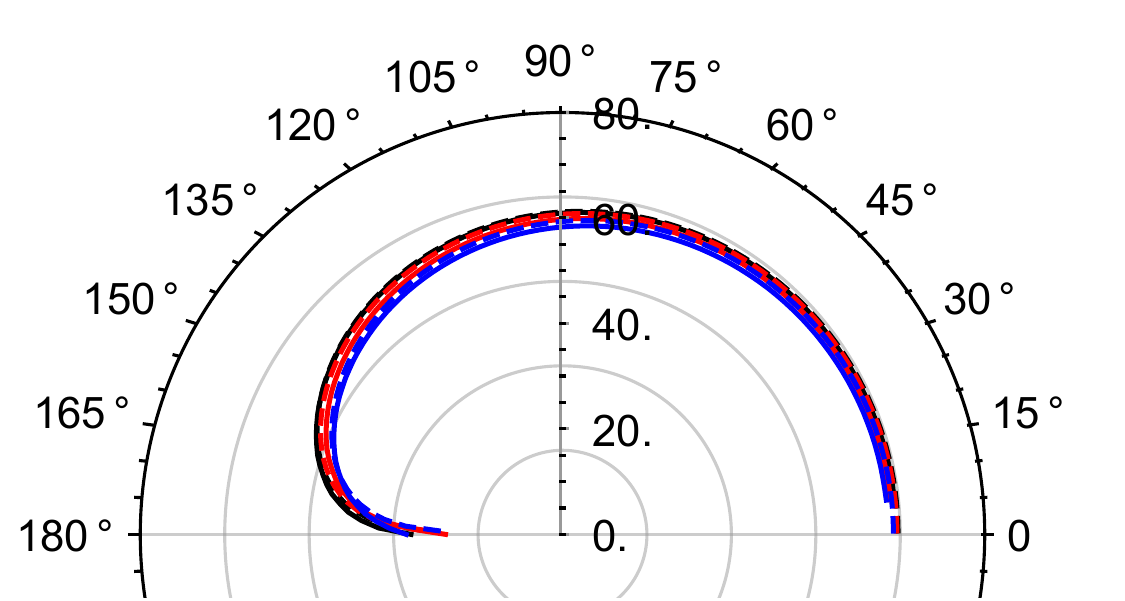}
\caption{$k_{1}=1$.}
\end{subfigure}
\hfill
\begin{subfigure}[b]{0.49\linewidth}
\includegraphics[width=1\textwidth]{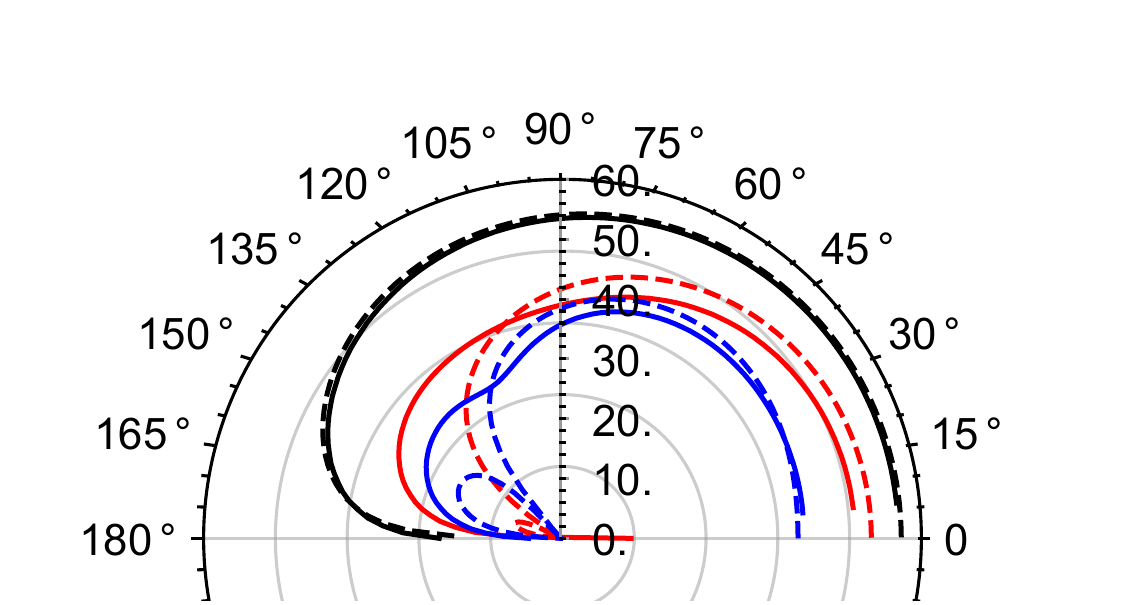}
\caption{$k_{1}=5$.}
\end{subfigure}
\\
\begin{subfigure}[b]{0.49\linewidth}
\includegraphics[width=1\textwidth]{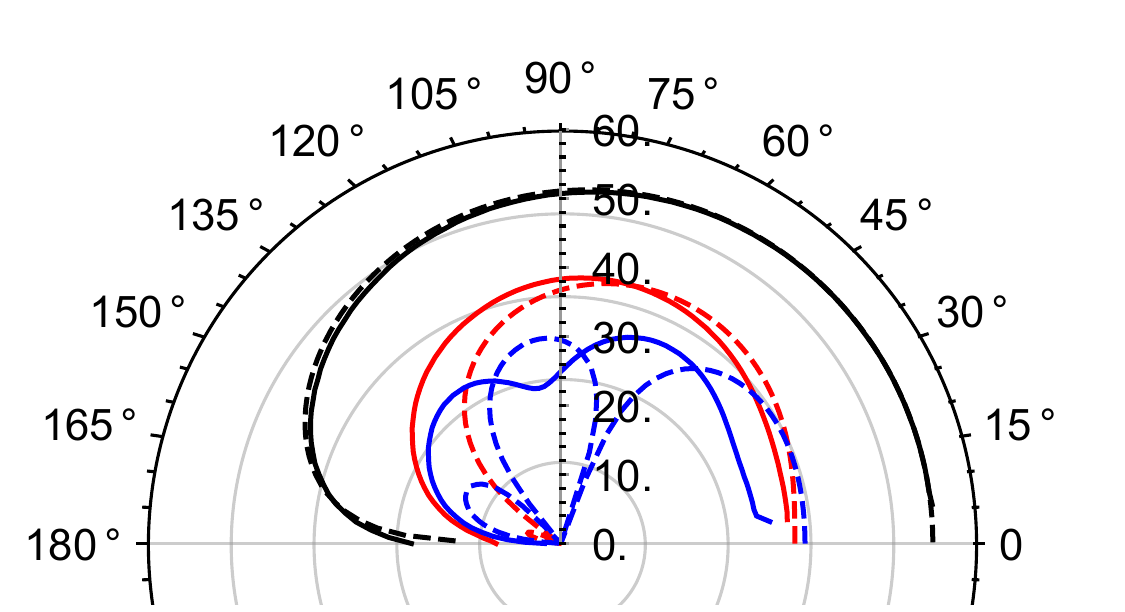}
\caption{$k_{1}=10$.}
\end{subfigure}
\caption{Polar plot of the (scaled) logarithmic spanwise average directivity, $10\log_{10}\left(10^{8}D_{n,a}(r,\theta)\right)$ calculated numerically (solid) and analytically (dashed), for different serration height parameters; $c=0$ (black), $c=2$ (red), $c=4$ (blue).}
\label{fig:directcompare}
\end{figure}

We next directly compare the numerical and analytical directivity patterns on a logarithmic scale by plotting $10\log_{10}\left(10^{8}D_{n,a}(r,\theta)\right)$ in Figure \ref{fig:directcompare}. The value of $10^{8}$ is included to ensure all plotted values are positive and visible. We see good agreement in the overall magnitude of the results, but some variation in the oscillations which we attribute to the differing incident fields generating different tip and root interference.

\begin{figure}
\centering
\includegraphics[width=0.5\textwidth]{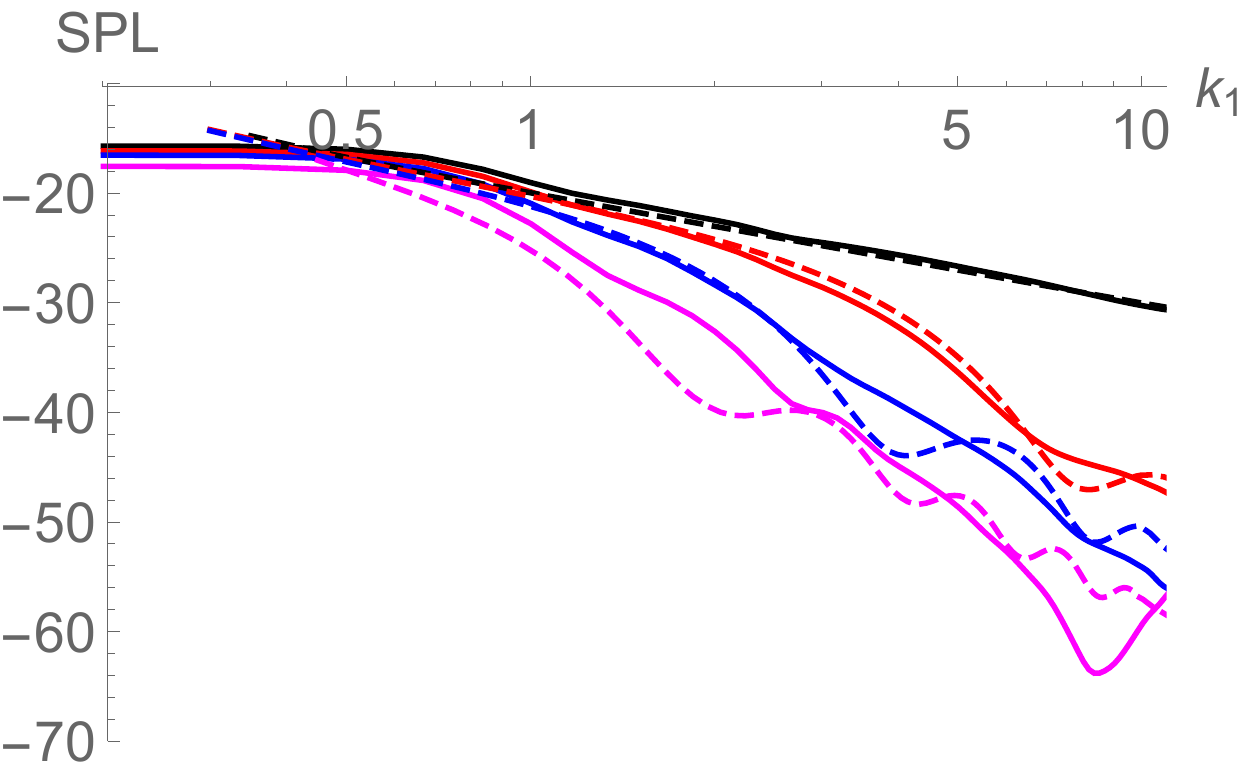}
\caption{Comparison of the $k_{3}=0$ SPL calculated numerically for a line vortex (solid) and analytically for an incident gust (dashed), for $c=0$ (black),  $c=2$ (red), $c=4$ (blue), and $c=8$ (magenta).}
\label{fig:SPLJW}
\end{figure}

To compare the two results over a wider range of frequencies, we now consider the far-field sound pressure level (SPL) measured in dB averaged over a cylinder at radius $r=10$, as a function of reduced frequency $k_{1}$. This is defined as
\begin{equation}
\text{SPL}=10\log_{10}\left(\frac{1}{\upi}\int_{0}^{\upi}\int_{0}^{1}|p(10,\theta,z)|^2 dz\,d\theta\right).
\end{equation}
We integrate over $\theta\in[0,\upi]$ to capture all oscillatory effects of the far-field directivity.
Analytically we restrict to $k_{3}=0$ (and thus do not include a turbulent spectrum) as the numerical line vortex cannot be given a spanwise dependency. We see very good agreement between the analytic and numerical results in Figure \ref{fig:SPLJW}. The very low frequencies do not agree well as the method of steepest descents required $k_{1}r\gg1$, and this assumption breaks down for small $k_{1}$. The highest $c$ results ($c=8$) compare least favourably, and we anticipate for such sharp serrations, the small distances between successive serrations enable stronger non-linear effects which are accounted for in the numerics but not the analytical model. As such the analytical model overpredicts the interference effect in the initial oscillation (for $k_{1}\approx2$). 

Overall we see that the two different turbulence models are capable of predicting the same levels of noise reduction for serrated leading edges.

\section{Conclusion}\label{sec:conc}
This paper has presented an analytic solution for the sound generated by a convective gust interacting with a flat plate with a serrated edge. The plate is in a channel with either rigid walls, or a periodic condition such that the result mimics that of a spanwise semi-infinite blade with leading-edge serration. The solutions follow the method used by \citet{Envia} who considered a swept blade in a channel with rigid walls and employed the Wiener-Hopf technique.

The solution both for rigid or periodic walls predicts overall a decrease in far-field acoustic pressure as the height of the serration in increased, however for low frequency incident gusts the reduction of noise is much smaller than at higher frequencies, as is known from experimental, numerical and analytical results \citep{Joseph1,Haeri,LyuLE}. There is little difference between the rigid or periodic cases when the gust has zero spanwise wavenumber, $k_{3}=0$, indicating the key parameters are the gust streamwise wavenumber, $k_{1}$, and the greatest distance between points along the leading edge, namely the tip-to-root distance, $c/2$. 
All of these features are also replicated in numerical results. The numerical method uses a different incident field, a line vortex, which is less suitable for a fully analytic solution (although a semi-analytic solution could likely be obtained if the incident field and Wiener-Hopf factorisation steps were performed numerically). Despite the differing incident fields there is good agreement for the far-field SPL verses frequency, $k_{1}$, for a range of serration heights, $c$. This indicates the key noise-reduction mechanisms are being appropriately modelled by the simple analytical solution which uses an incident gust, and the numerical scheme which uses an incident line vortex. 

% The analytical solution also permits an incident gust with a non-zero spanwise wavenumber component. In such a case there is the possibility of noise increases for a serrated edge versus a straight edge, particularly for periodic wall conditions. However, for periodic wall conditions, a non-zero $k_{3}$ can also cut off all acoustic modes in a similar manner to a spanwise infinite swept edge (for a swept edge in a channel with rigid walls $k_{3}$ does not act to cut off noise).

The analytical solution relies on a modal decomposition of the incident and scattered fields, however this is not simply a Fourier series expansion, but a modal expansion which depends intrinsically on the leading-edge geometry. Because of this choice of modal basis, the expansion coefficients can be calculated analytically thus the acoustic pressure can be computed for any parameters incredibly quickly; span-averaged directivities can be plotted in approximately 6 seconds on a standard 4-core desktop computer (via Mathematica).
The modal coefficients also shed light on the mechanisms behind the noise reduction as we see they lead to two key features. First, the coefficients are oscillatory indicating interference in the acoustic pressure in the far field. This interference commonly is destructive and reduces the scattered noise, but in some cases of non-zero $k_{3}$ has been seen to increase the scattered noise (i.e. is constructive). Second, as the tip-to-root distance of the serration, $c/2$, increases, the expansion coefficients vary, with lower modes tending to zero. Higher modes do not tend to zero but are cutoff. Thus increasing the tip-to-root distance redistributes acoustic energy from the lower (cuton) modes to higher (cutoff) modes hence significant far-field noise reductions can be achieved for large tip-to-root distances. 
This second noise reduction mechanism means that even if a constructive interference were to occur, an overall noise reduction would be observed in the far-field.

The rate at which the cuton modes decrease with increasing serration height is proportional to $1/c$, thus it is predicted as $c$ increases the reduction of noise in decibels is logarithmic, $\sim\log_{10}(c)$, as alluded to in \citet{sn15}. A logarithmic noise reduction dependency indicates that continuing to increase the tip-to-root serration height lessens the level of noise reduction. Therefore, since increasing the tip-to-root height decreases aerodynamic performance, for a given application of serrated leading edges, it is likely an optimum serration height yielding a significant noise reduction could be determined for a prescribed limited decrease of aerodynamic performance. However obtaining further noise reductions, whilst possible by further increasing the tip-to-root height, would be too costly on aerodynamic performance. Other leading-edge designs, such as the hook structures from \citep{Joseph1} (comprised of a sawtooth serration with an additional v-shaped cut in the root), could allow the same tip-to-root ratio as a simple serration, but greater aerodynamic efficiency due to a greater leading-edge surface area. Therefore much work is still needed to consider more complicated leading-edge geometries and the potential for optimal noise reduction with minimal aerodynamic impact.

\section*{Acknowledgements}
The authors would like to acknowledge the support from EPSRC Fellowship EP/P015980/1 (L.A.), and from the IRIDIS-4 supercomputing facility and services provided by the University of Southampton (J.W.K.) for the completion of the work.

\appendix
\numberwithin{equation}{section}
\section{Expansion Coefficients, $E_{n}(\lambda)$}\label{app:1}
The expansion coefficients are obtained by
\begin{equation}
 E_{n}=\epsilon_{n}^{-1}\int_{0}^{1}\e^{i\kappa\gamma F(\zeta)+i k_{3}\zeta}\overline{Z_{n}(\bar{\lambda},\zeta)}d\zeta,
\end{equation}
where the overbar denotes the complex conjugate, and $\epsilon_{n}$ are normalisation coefficients
\begin{equation}
 \epsilon_{n}=\int_{0}^{1}Z_{n}(\lambda,\zeta)\overline{Z_{n}(\bar{\lambda},\zeta)}d\zeta.
\end{equation}

When $Z_{n}$ is given by \eqref{eq:Zwalls} (case a), 
\begin{equation}
 E_{0}=\frac{2\e^{i k_{3}/2}}{(k_{3}-s)(k_{3}+s)}\left(\left(k_{3}-s\right)\sin\left[\frac{k_{3}}{2}\right]+2s\,\sin\left[\frac{1}{4}\left(k_{3}-s\right)\right]\right),
\end{equation}
\begin{align}
 E_{n}=&\frac{\e^{-\frac{i s}{4}}}{((k_{3}+s)^{2}-n^{2}\upi^{2})((k_{3}-s)^{2}-n^{2}\upi^{2})}\left[
2\e^{\frac{i s}{4}}\left(
1-(-1)^{n}\e^{i k_{3}}
\right)
(k_{3}+s)((k_{3}-s)^{2}-n^{2}\upi^{2})\right.\notag
\\
&+4i s(k_{3}^{2}-s^{2}+n^{2}\upi^{2})\cos\left[
\frac{n\upi}{4}
\right]
\left(
\e^{\frac{i}{4}(k_{3}+2s)}-(-1)^{n}\e^{\frac{3i k_{3}}{4}}
\right)\notag
\\
&\left.+8k_{3}n\upi s\sin\left[
\frac{n\upi}{4}
\right]
\left(
\e^{\frac{i}{4}(k_{3}+2s)}+(-1)^{n}\e^{\frac{3i k_{3}}{4}}
\right)
\right],
\end{align}
where $s=\gamma(\kappa+\lambda)$.
We note that all singularities in these $E_{n}$ are removable, thus there are no poles.

When $Z_{n}$ is given by \eqref{eq:Zperiodic} (case b), 
\begin{align}
 E_{0}&=\frac{2\e^{i q/2}}{s^{2}-q^{2}}\left(2 s\sin\left[\frac{s-q}{4}\right]+(s-q)\sin\left[\frac{q}{2}\right]\right),
\\
E_{n}&=\frac{-i\e^{-i s/4}}{\left(s+2n\upi-q\right)\left(s-2n\upi+q\right)}\left(2i^{n}s \e^{3i q/4}\left(-1+\e^{i(s+2n\upi -q)/2}\right)+\e^{i s/4}(e^{i q}-1)(s+2 n\upi-q)\right),
\end{align}
where $q=k_{3}(1+\alpha)$.

 \bibliographystyle{jfm}

\end{document}